%% file: sample-acmsmall.tex
\documentclass[manuscript]{acmart}

\usepackage{color, colortbl}
\definecolor{LightCyan}{rgb}{0.88,0.88,0.88}
\usepackage{graphicx}
\usepackage{multirow}
\usepackage{threeparttable} 
\usepackage{pdflscape}

\usepackage{bbding}
\begin{document}

\title{Towards Human-centered Design of Explainable Artificial Intelligence (XAI): A Survey of Empirical Studies}


\author{Shuai Ma}
\orcid{0000-0002-7658-292X}
\affiliation{
  \institution{The Hong Kong University of Science and Technology}
  \city{Hong Kong}
  \country{China}
}
\email{shuai.ma@connect.ust.hk}

\renewcommand{\shortauthors}{}

\begin{abstract}
With the advances of AI research, AI has been increasingly adopted in numerous domains, ranging from low-stakes daily tasks such as movie recommendations to high-stakes tasks such as medicine, and criminal justice decision-making. Explainability is becoming an essential requirement for people to understand, trust and adopt AI applications.

Despite a vast collection of explainable AI (XAI) algorithms produced by the AI research community, successful examples of XAI are still relatively scarce in real-world AI applications. This can be due to the gap between what the XAI is designed for and how the XAI is actually perceived by end-users. As explainability is an inherently human-centered property, in recent years, the XAI field is starting to embrace human-centered approaches and increasingly realizing the importance of empirical studies of XAI design by involving human subjects.

To move a step towards a systematic review of empirical study for human-centered XAI design, in this survey, we first brief the technical landscape of commonly used XAI algorithms in existing empirical studies. Then we analyze the diverse stakeholders and needs-finding approaches. Next, we provide an overview of the design space explored in the current human-centered XAI design. Further, we summarize the evaluation metrics based on evaluation goals. Afterward, we analyze the common findings and pitfalls derived from existing studies. For each chapter, we provide a summary of current challenges and research opportunities. Finally, we conclude the survey with a framework for human-centered XAI design with empirical studies.

\textbf{Author's Note: This manuscript was written in 2022 May, so the surveyed literature is not up-to-date. During the writing, I refereed a lot from Vivian Lai's paper \cite{lai2021towards} and Vera Liao's paper \cite{liao2021human}. Since May 2022, many empirical studies on XAI have been published. Nevertheless, I hope this manuscript can serve as a starting point for interested readers.}

\end{abstract}





\maketitle
\newpage
\tableofcontents
\newpage
\input{sections/introduction}

\input{sections/techniques}

\input{sections/userNeeds}

\input{sections/designs}

\input{sections/evaluations}

\input{sections/findings}
\input{sections/framework}

\bibliographystyle{ACM-Reference-Format}
\bibliography{sample-base}

\end{document}

%% file: sections/introduction.tex
\section{Introduction}

% Artificial intelligence (AI) has increasingly penetrated everyday applications \cite{eiband2021support, yang2020re}. In recent years, the HCI and AI communities have had extensive discussions around the challenge of the opaqueness of intelligent systems \cite{rader2018explanations, cheng2019explaining, liao2020questioning, benjamin2021machine}.

With the advances of AI research, AI has been increasingly introduced into different tasks in people’s work and life, ranging from low-stakes daily tasks such as movie recommendations, virtual assistant agents, to high-stakes tasks such as medicine, criminal justice, etc \cite{annappindi2014system, dastin2018amazon, dilsizian2014artificial, khandani2010consumer, wang2014improving, ma2019smarteye, yang2018insurance}. When users interact with a system, users' ``mental model'' (an understanding of how the system works) \cite{norman2013design}, plays a fundamental role for users to correctly and effectively interact with the system. However, many AI models are hard for users to understand due to their black-box nature \cite{adadi2018peeking}. This lack of transparency can lead to users' inappropriate mental models of how the model works, which can cause other problems such as inappropriate trust and unexpected adoption of the intelligent system \cite{bussone2015role}. Besides, a lack of explainability of AI might cause failures in usability and moral crises, such as fairness, reliability, safety, accountability, etc \cite{liao2021human}. Furthermore, some legal requirements have been proposed \cite{arrieta2020explainable, adadi2018peeking}, such as GDPR's requirement that AI applications must provide people who are affected by automated systems with ``meaningful information about the logic involved''. 

To solve these problems, a lot of eXplainable Artificial Intelligence (XAI) algorithms have been introduced, where the AI model explains its reasoning process to the users \cite{mohseni2021multidisciplinary}. Explanations can be any information that is beneficial for users to understand the AI model, ranging from logical information about the model, reasons for a prediction, to general information about the training data, input/output space, and more \cite{liao2020questioning, liao2021human, arrieta2020explainable}.

Many works in the XAI domain make technical contributions, such as developing new algorithms that increase the interpretability of the AI model as well as ensuring the model's performance \cite{adadi2018peeking}. Meanwhile, the XAI community is increasingly realizing the importance of empirical studies of XAI design by involving human subjects. On the one hand, to design an effective and appropriate explanation, designers need to understand users' specific needs in the target task based on user research \cite{liao2021human}. On the other hand, the effectiveness of explanation is determined by the perception and reception of the explainee (receiver of an explanation, i.e., the human) \cite{ehsan2021explainable} because they own different knowledge, expertise, background, role, needs, and goal for receiving the explanation \cite{liao2021human}.

The goals of XAI empirical studies are not only to evaluate the effectiveness of a specific XAI design, but also for researchers and designers to better understand how target users actually interact with the designed XAI to solve real-world tasks. With a comprehensive understanding of users' perceptions and utilization of the explanatory interfaces, empirical studies can contribute on multiple grounds, including: (1) to guide new explainable techniques that provide more interpretable and effective explanations; (2) to provide insights for designers to design more effective explanation for designated human-AI interaction scenarios; (3) to understand how to integrate the knowledge of human's cognitive processes and mental models, such as theories from psychology, cognitive science, social science, into human-centered XAI design.

Recently, a research community of human-centered XAI \cite{ehsan2020human, ehsan2021operationalizing, liao2021human, wang2019designing} has emerged. However, there is a lack of systematic investigation of human-centered XAI design from the empirical study perspective. This hinders the emergence of systematic knowledge and joint research effort \cite{lai2021towards}. We recognize several challenges. First, existing XAI empirical studies are built on different XAI techniques and conducted with different stakeholders for different explainability needs. Without systematically reviewing the scope of these factors, we may not be able to form a whole picture of current XAI studies. Second, based on different design goals and research questions, different empirical studies tend to focus on various XAI designs. There is a lack of overview of the large design space, i.e., how different explanatory interactions are designed in existing empirical studies. Third, the evaluation method of empirical studies is varying for different research questions, evaluation goals, etc. This leads to the fact that the effects of XAI design are measured by different evaluation methods in different studies, even though the design goals are similar. In other words, XAI design lacks a unified evaluation system. Lastly, existing empirical studies have identified a rich set of empirical findings which can help researchers and practitioners in the XAI domain understand the nuanced effects of different explanation designs. However, there lacks an organized summarization of these off-the-shelf findings, hindering knowledge sharing and transfer in this domain.

To move the first step towards a systematic review of empirical study for human-centered XAI design, we systematically investigate the current state of the field. We review works focusing on exploring human needs of XAI through user research, evaluating the effects of the proposed XAI design with empirical studies, and understanding humans’ perceptions and experiences when interacting with specific XAI designs, rather than works focusing on developing new XAI algorithms or developing a system without any empirical studies. The scope of this survey differentiates from either prior surveys on XAI that deviate from the focus of empirical study \cite{ehsan2020human, ehsan2021operationalizing, liao2021human, sperrle2020should} or empirical studies of human-AI interactions that do not focus on XAI \cite{sperrle2021survey, vereschak2021evaluate, amershi2019guidelines}. To mitigate the four above-mentioned challenges, our survey focuses on analyzing four aspects of empirical study in these surveyed papers: the stakeholders and needs, the design space explored in existing work, the evaluation metrics of different design goals, and the common findings and pitfalls from existing empirical studies.

The remainder of this survey is structured as follows. We first start with a brief overview of XAI techniques commonly adopted in existing empirical studies. Then we analyze the diverse stakeholders and need-findings approaches. Next, we show an overview of the current explored XAI design space. Further, we summarize the evaluation metrics based on the evaluation goals. Afterward, we analyze the common findings and pitfalls derived from existing studies. Finally, we conclude the survey with a framework for designing human-centered XAI with empirical studies.

% call to actions for establishing a framework to integrating empirical studies, grounded theories, and technical development together for this emerging area.

%% file: sections/techniques.tex
\section{Commonly Adopted XAI Techniques in Empirical Studies}

There are some relevant terms with ``explainability'', such as ``interpretability'', ``intelligibility'', ``transparency'' \cite{mohseni2021multidisciplinary}. In this survey, we adopt the most commonly used term ``explainability'' which shares a goal to make AI understandable to users. Recent papers have surveyed this field from a technical perspective and converged on several important dimensions to classify XAI algorithms \cite{liao2021question, arrieta2020explainable, arya2019one, adadi2018peeking}. One categorization is based on the intrinsic transparency of the AI models, which can be divided into directly explainable models (such as rule-based models, linear model, decision trees) and opaque models (such as deep neural networks), and the latter often require using additional algorithms to generate post-hoc explanations \cite{guidotti2018survey, lipton2018mythos}. Another dimension widely used in XAI classification is the scope of explanation. XAI techniques can be categorized into \emph{global} explanation of the overall logic of the model, and \emph{local} explanation of a specific prediction \cite{liao2021human}. In this chapter, we also use this dimension to categorize and conclude common explanation methods adopted in existing empirical studies for XAI, and describe each briefly. Note that we will treat all explanatory approaches as XAI methods as long as they can benefit users in understanding the AI. So, apart from model inner working-related explanation and feature-related explanation, we also include model-related factors into our XAI methods coverage, such as training data-related explanation, uncertainty-related explanation, performance-related explanation, etc.

\renewcommand{\arraystretch}{1.5}
\begin{table}[tp]  
  \centering  
  \fontsize{8}{8}\selectfont

\begin{threeparttable}  

\caption{XAI Techniques Category Based on Explanation Scope.}\label{table:posttask-questionnaire-student} 
\begin{tabular}{p{1cm}p{4cm}p{9cm}}
%\begin{tabular}{cccccc}

\toprule
\textbf{Scope}&\textbf{Category}&\textbf{Method and Example}\\ \hline
\multirow{8}*{Global}&Global feature importance&Permutation-based \cite{fisher2019all, wang2021explanations}, coefficients \cite{dodge2019explaining}\\\cline{2-3}

&Global example-based explanations&Model tutorial \cite{lai2020chicago}, prototypes \cite{nguyen2016synthesizing, cai2019effects, feng2019can}\\\cline{2-3}

&Presentation of simple models&Decision trees \cite{slack2019assessing}, linear regression \cite{poursabzi2021manipulating}, decision sets \cite{lakkaraju2016interpretable}, logistic regression \cite{slack2019assessing}, one-layer MLP \cite{slack2019assessing}\\\cline{2-3}

&Model performance&Accuracy \cite{harrison2020empirical, lai2020chicago, lai2019human, yang2020visual, yin2019understanding}, false positive rates \cite{harrison2020empirical}\\\cline{2-3}

&Model documentation&Overview of the model or algorithm \cite{kocielnik2019will, kulesza2012tell, kulesza2013too, lee2019procedural, rader2018explanations}, model prediction distribution \cite{van2021effect}\\\cline{2-3}

&Information about training data&Input features or information the model considers \cite{dodge2019explaining, harrison2020empirical, poursabzi2021manipulating, zhang2020effect}, aggregate statistics (e.g., demographic) \cite{binns2018s, dodge2019explaining}, full training “data explanation” \cite{anik2021data}\\\hline

\multirow{8}*{Local}&Model uncertainty&Classification confidence (or probability)  \cite{arshad2015investigating, bansal2021does, buccinca2021trust, bussone2015role, feng2019can, guo2019visualizing, lin2020limits, lee2020co, zhang2020effect}\\\cline{2-3}

&Example-based methods&Nearest neighbor or similar training instances \cite{buccinca2020proxy, cai2019effects, cai2019human, dodge2019explaining, hase2020evaluating, lai2019human, tsai2021exploring, wang2021explanations}\\\cline{2-3}

&Local feature importance&Coefficients \cite{cheng2019explaining, dodge2019explaining, ghai2021explainable, green2019principles, lai2020chicago, lai2019human, liu2021understanding, poursabzi2021manipulating}, attention \cite{carton2020feature, chandrasekaran2018explanations, lai2020chicago}, gradient-based \cite{chandrasekaran2018explanations, kiani2020impact, nguyen2018comparing}, propagation-based (LRP \cite{alqaraawi2020evaluating}), perturbation-based (LIME \cite{alqaraawi2020evaluating, hase2020evaluating, nguyen2018comparing}, SHAP \cite{weerts2019human, zhang2020effect})\\\cline{2-3}

&Rule-based explanations&Decision sets \cite{kulesza2013too}, tree-based explanation \cite{lage2019evaluation}, anchors \cite{ribeiro2018anchors}\\\cline{2-3}

&Counterfactual/Contrastive explanations&Contrastive or sensitive features \cite{dodge2019explaining, lucic2020does}, counterfactual examples \cite{slack2019assessing, wang2021explanations}\\

\bottomrule
\end{tabular}
\end{threeparttable}  

\end{table}

\subsection{Global Explanation.}

Stakeholders of AI applications often need to understand the underlying logic of an AI model to form an accurate mental model. ``Global'' explanation about the model can include global feature importance, global example-based explanations, presentation of simple models, model's overall performance, model documentation, information about the training data, etc.

\subsubsection{Global feature importance}
Global feature importance quantifies the overall importance of each feature used to get the model’s decisions. Some models can directly produce feature importance, such as coefficients in linear or logistic regression models \cite{dodge2019explaining} and shape function of GAMs \cite{abdul2020cogam}. Another way to get feature importance is from the post-hoc model, such as permutation importance \cite{fisher2019all, wang2021explanations}. Note that feature-importance methods can also be used for local explanation.

\subsubsection{Global example-based explanations}
Example-based explanations provide data examples for users to understand how the model works. One way is to select some instances from the training set that can provide insights to the user. For example, Lai et al. \cite{lai2020chicago} select examples from the training set as a tutorial. Another common approach is that for a given prediction class, selecting one or a set of training instances that are representative and have the same class labels \cite{nguyen2016synthesizing}. The example-based explanation can also be used to locally explain a prediction. The main difference is that the representative instances are often provided in the onboarding stage for the global explanation.

\subsubsection{Presentation of simple models}
For a complex black-box model, it is hard for users to understand its complex internals. Usually, post-hoc XAI methods are utilized to train a simple directly interpretable model such as a decision tree, rule set, with the same training data to provide an approximate overview of how the complex model behaves \cite{liao2021human, tan2018learning, dhurandhar2020enhancing}. For a simple model, we can directly present the model internals to users.
% For this reason, we can find that in existing empirical studies, more papers choose simple models when interpretability is desired.

\subsubsection{Model performance}
Model performance can provide the basic information of how well the model works in general to uses. In the empirical studies with classification tasks, model performance is mainly presented in the form of accuracy \cite{harrison2020empirical, lai2020chicago, lai2019human, yang2020visual, yin2019understanding}. These works typically explore the effects of whether to show or how to show the model performance on users' perception of and decision making with the model \cite{lai2019human, yin2019understanding}. Note that accuracy is usually estimated on the validation set, and the model’s actual performance can be different from the estimated performance because the model has to perform prediction on unseen data.

\subsubsection{Model documentation.}
Some literature designs model documentation to provide the meta information of a model. This meta information includes the characteristics of the model, how and for what purpose the model is developed, which is verified to be critical to AI transparency \cite{lai2021towards, arnold2019factsheets, mitchell2019model}. For example, some ``About Me'' page Model cards have developed \cite{arnold2019factsheets, mitchell2019model}. Kocielnik et al. \cite{kocielnik2019will} propose a meeting scheduling assistant that displays a description of how it works.

\subsubsection{Information about training data.} 
As the training data plays a critical role in the model development, providing the information about the training data, such as data distribution and feature set, is verified to be able to help users better understand the model \cite{dodge2019explaining, harrison2020empirical, poursabzi2021manipulating, zhang2020effect}. For example, Zhang et al. \cite{zhang2020effect} investigate whether the additional knowledge can affect users' trust in a simulated income prediction task, where the participants are made aware of whether or not the model has considered ``marital status'' as a feature.
% Finally, Anik and Bunt \cite{anik2021data}, in a more detailed way, presents a ‘full’ training data explanation, including how the data was collected, demographics, recommended usage, potential issues.

\subsection{Local Explanation}
Local explanations are usually used to explain how a specific prediction is made by the model. The explanation information of a local prediction includes local feature importance, example-based explanations, rule-based explanations, counterfactual explanations, and model’s uncertainty for the prediction, etc.

\subsubsection{Model uncertainty.}
Although a model's uncertainty or confidence for a prediction is not a direct explanation for how and why the prediction is made, it can bring valuable information about how confident the model is for the prediction. Then the users can decide whether to rely on the model based on the uncertainty \cite{ghosh2021uncertainty}. The uncertainty or confidence is usually calculated as the probability of the predicted label in a classification task, ranging from 0 to 1 (or scaled to 0 to 100). A lot of empirical studies have investigated the effects of showing model uncertainty on users \cite{arshad2015investigating, bansal2021does, buccinca2021trust, bussone2015role, feng2019can, zhang2020effect}.

\subsubsection{Example-based methods.}
Example-based methods use similar examples to the current instance to support case-based reasoning. Generally, for a target instance to explain, the model will research similar examples from the training data which have the same label/class as the target instance. The similar examples are usually found by some similarity/distance metrics in the embedding space or feature space. The example-based explanation had been widely adopted in many empirical studies \cite{buccinca2020proxy, cai2019effects, cai2019human, dodge2019explaining, lai2019human, tsai2021exploring, wang2021explanations}.

\subsubsection{Local feature importance.}
The local feature importance explanation will calculate the importance of each feature to the current prediction. For example, when predicting income, some features play a more important role to the prediction (e.g., the occupation) while others might be less important (e.g., the weight of a person). Generally, the local feature importance can be obtained in two ways, the built-in method and the post-hoc methods.

Built-in methods can be mainly categorized into the coefficient-based method and attention-based method. First, some models can generate coefficients as feature importance. For instance, the coefficient of a linear regression model can be seen as a direct measure of feature importance. Coefficient-based feature importance has been adopted in many empirical studies \cite{cheng2019explaining, dodge2019explaining, ghai2021explainable, green2019principles, lai2020chicago, lai2019human, liu2021understanding, poursabzi2021manipulating}. Second, the attention-based method is often used for explaining deep learning models, such as attention in an NLP model and saliency region in a computer vision model. This method has also been verified in some works \cite{carton2020feature, chandrasekaran2018explanations, lai2020chicago}. For example, Lai et al. \cite{lai2020chicago} compare the attention-based explanation with a LIME-based explanation.

Post-hoc methods usually train a separate model for the original model, often used for black-box models. And the feature importance is generated from the new model. Post-hoc methods can be categorized into gradient-based \cite{chandrasekaran2018explanations, kiani2020impact, nguyen2018comparing}, perturbation-based \cite{alqaraawi2020evaluating, hase2020evaluating, chromik2021think, zhang2020effect} and propagation-based \cite{alqaraawi2020evaluating}. From the existing studies, we find that LIME \cite{alqaraawi2020evaluating, hase2020evaluating} and SHAP \cite{weerts2019human, zhang2020effect} are two widely used methods.

\subsubsection{Rule-based explanations.}
Rule-based methods, such as a set of ``if-then'' explanations can be easily understood by humans. Similar to the feature importance explanation, the rule can be directly got from simple models, such as decision trees \cite{lage2019evaluation}, decision sets \cite{kulesza2013too}. Also, the rule can be generated from post-hoc methods, such as anchors \cite{ribeiro2018anchors}. Compared to example-based and feature importance-based explanations, the rule-based explanation is less investigated in existing empirical studies.

\subsubsection{Counterfactual/Contrastive Explanations.}

Counterfactual explanations are widely used when users would like to know how the prediction will change if the current input changes. If a user gets a prediction that is not her expected, she might be interested in figuring out ``how to change to get a different prediction'' or ``why not a different prediction''. For example, a loan declined user wants to know how to get approved. In some surveys \cite{liao2021human, liao2021question}, counterfactual methods are differentiated from local explanation because it is not the direct explanation for a prediction. However, in this paper, we categorize it into the local scope as it is often desired after a user sees a specific prediction.

In the research literature, contrastive explanations are often conflated with counterfactual explanations \cite{zhang2021towards}. Although similar, they have differences. Generally, contrastive explanations are used to answer ``Why Not'' questions, and counterfactual explanations focus on answering ``How To Be That'' questions \cite{lim2009assessing}.

\subsection{Summary and Thinking}
There have been a rich amount of XAI techniques which offer opportunities to design explanation for both simple models and complex black-box models. Categorized by the explanation scope, these techniques can be divided into global methods and local methods. XAI designers can select appropriate XAI techniques based on the type of AI models and the specific explanation goals. We also find some gaps.

\subsubsection{Challenges}

\textbf{Limited AI types and task types applied in empirical studies.}
From the surveyed papers, we find that most studies use simple models for interpretability. While for complex models, such as DNN, CNN, GNN, and RL, there is still a lack of empirical studies for non-expert users, although there are corresponding explanation technologies and some visual interaction systems have been developed to help model developers understand and debug models. This needs to solve several problems, first of all, although complex models can be approximated as simple models or explained by posthoc methods, the faithfulness issue should be considered. Second, the complex information brought by the explanation of complex models requires more expertise from ordinary users. Therefore, it is necessary to seek a balance between the accuracy and comprehensiveness of the explanation and the ease of interpretation.

Another trend is that in most of the current empirical studies, simplified experimental tasks are selected, usually a binary classification task is used as the testbed. However, tasks in the real world are not only binary classification, such as multi-classification or continuous prediction tasks (regression tasks). Although there are several works that choose regression tasks, such as apartment price prediction \cite{poursabzi2021manipulating}, research in this area is relatively rare.

\subsubsection{Future research opportunities}
\textbf{Informing new XAI techniques by human-centered research.} One way that human-centered research can contribute is to offer more implications to the technical development of XAI. One representative example is RexNet published at CHI2022 \cite{zhang2021towards}, which generates relatable explanations inspired by humans' perceptual process from cognitive psychology. There are rich cognition, psychology, and social theories under exploration in XAI algorithm design.

\textbf{Investigating the effects of more XAI algorithms in more diverse tasks.} In the future, explanations for other types of AI should be explored in empirical studies, such as Reinforcement Learning models. And HCI researchers can pay more attention to regression tasks which are under explored.

% A small but growing number of prior works studying human-AI decision making have utilized counterfactual explanations, such as contrastive and sensitive features [39, 97]. For example, Dodge et al. [39] present sensitivity-based explanations to help people discover unfair decisions (by changing just the race feature, the model would have a different prediction). And, Lucic et al. [97] generate contrastive explanations for model errors by identifying feature ranges that lead to reasonable predictions. Similarly, other work uses counterfactual examples include [15, 45, 92, 143]. For example, Wang and Yin [143] show instances with minimal changes that result in the desired output.

% \subsubsection{Natural language explanations.}

% Natural language explanations, or sometimes referred to as rationale-based explanations, are a form of “why” explanations that provide the reasoning or rationale behind a particular decision. For example, Tsai et al. \cite{tsai2021exploring} study rationale-based explanations for their COVID-19 chatbot, such as why the chatbot asks particular diagnostic questions to the user. These explanation types are sometimes referred to as “justifications” \cite{biran2017human}. Natural language explanations can be differentiated by how they are generated, either model/algorithm generated \cite{biran2017human}[16, 32, 137]— where these explanations are produced by the system—or human experts generated [13], meaning domain experts (or algorithm developers) provided rationales behind types of predictions to be shown to users.

%% file: sections/userNeeds.tex
\section{Different Stakeholders and Various Explainability Needs}
Different stakeholders of AI application have different needs for explanation \cite{liao2020questioning, mohseni2021multidisciplinary, liao2021human}. To understand users' explainability needs, some papers categorize users into different groups, while another lines of papers adopt human-centered user research to derive user needs.

\subsection{User Group-Driven Explainability Needs}
There are mainly two types of user group categorization methods. One is based on users' expertise and knowledge, and the other is based on users' role when interacting with AI applications.

\subsubsection{Expertise-Based User Group Categorization and Corresponding Explainability Needs}

User expertise/knowledge is one commonly recognized characterization. Literature has shown that the XAI design goals can be different for different levels of expertise. For example, Mohseni et al. \cite{mohseni2021multidisciplinary} categorize users into ``AI experts'', ``data experts'', and ``AI novices''. The design goals for AI experts can include model interpretability and model debugging. The design goals include model visualization and inspection, and model tuning and selection for data expert (also called domain expert in other papers, who is an expert in the AI application domain but has little knowledge of AI). And the design goals for AI novices can include AI transparency, user trust, bias mitigation, and privacy awareness. Recently, Suresh et al. \cite{suresh2021beyond} introduce a framework with a more granular characterization for the stakeholders based on their knowledge and objectives. From the expertise perspective, they characterize stakeholders' knowledge (i.e., formal, instrumental, and personal knowledge) and the context to which the knowledge belongs (i.e., machine learning, the data domain, and the milieu). From the objective perspective, they identify three-level of goals, from long-term goals to shorter-term objectives to immediate goals.

\subsubsection{Role-Based User Group Categorization and Corresponding Explainability Needs}

Another commonly recognized characterization of users is based on users' functional roles. For example, Arrieta et al. \cite{arrieta2020explainable}, Hind et al. \cite{hind2019explaining} and Preece et al. \cite{preece2018stakeholders} summarize several common user groups and map the corresponding needs, including (1) Model developers whose needs are to debug or improve a model; (2) Business owners or administrators whose needs are to assess an AI application’s regulatory compliance and capability; (3) Decision-makers whose needs are to make informed decisions with appropriate trust; (4) Impacted groups, whose needs are to seek recourse or contest the AI; (5) Regulatory bodies whose needs are to audit for legal or ethical compliance such as fairness, safety, privacy, etc. Based on such categorization, the role a person plays during the human-AI interaction can determine their explainability needs.

\subsection{User Research-Driven Explainability Needs}
In the XAI domain, researchers have increasingly realized that a good XAI design needs to understand the actual needs of users. One of the most direct and effective methods might be to obtain first-hand user needs based on user research methods such as interviews, surveys, formative studies, participatory design, etc. For example, for the explanatory interface design of an AI-assisted admission prediction system, Cheng et al. \cite{cheng2019explaining} organize a workshop inviting several types of stakeholders, including algorithm experts, UI designers, prospective students interested in applying to the university, current graduate students, faculty members to join the workshop. For a COVID-19 symptom checker, Tsai et al. \cite{tsai2021exploring} conduct a first user study to investigate what kinds of explanations users really need. To help clinical decision-making, Cheng et al. \cite{cheng2021vbridge} develop VBridge. To figure out users' needs for explainable AI, they conduct a pilot study with six clinicians to understand their expected functions and explanation methods.

One representative work in recent years is the question-driven explainability needs categorization proposed by Liao et al. \cite{liao2020questioning, liao2021question}. In a work called questioning the AI \cite{liao2020questioning}, they propose to identify users' explainability needs by figuring out what questions users want to ask. They contribute an \emph{XAI Question Bank}, with more than 50 detailed user questions organized in 9 categories, including \emph{How}, \emph{Why}, \emph{Why Not}, \emph{How to be That}, \emph{How to Still Be This}, \emph{What if}, \emph{Performance}, \emph{Data}, \emph{Output}. Each category of questions is related to some types of explanations. This XAI question bank can be used as a tool to identify applicable questions in user research. In a follow-up work \cite{liao2021question}, they propose a question-driven user-centered design method. First, designers can identify questions users might ask through user research. Then, designers can choose XAI techniques and iteratively design the interface based on questions. It is worth mentioning that they map the XAI technique space (as well as corresponding open-source XAI toolkits) with the user question categories, which is very helpful to select specific XAI methods for a specific user question to solve.

\subsection{Summary and Thinking}

There is a problem with the \emph{User Group-Driven} explainability needs acquisition because beyond users' expertise and knowledge, a large scope of user characteristics can affect how an explanation works for a specific user, such as users' locus of control \cite{rotter1966generalized}, need for cognition \cite{cacioppo1984efficient}, visual literacy \cite{abdul2020cogam, boy2014principled}, etc. Due to the diversity of users, there are always complex intersections between different user groups, and even users belonging to the same group are still different in many other perspectives, and these differences can affect their explainability needs and the effects of XAI on them. We believe that the \emph{User Group-Driven} method can play a guiding role in the design of XAI, and the two methods should be combined. Specifically, \emph{User Group-Driven} method can be used as a benchmark to guide the direction of user research, such as the questions design in a pilot study. At the same time, in order not to be limited by the division of fixed user groups, when designing XAI in a new scenario, it is necessary to make full use of \emph{User Research-Driven} method to obtain specific user needs, so as to better match it to specific task scenarios. There are already some good practices. For example, Cai et al. \cite{cai2019hello} and Tonekaboni et al. \cite{tonekaboni2019clinicians} follow this approach of role-based needs-finding as well by interviewing clinicians. However, we identify some gaps.
\subsubsection{Challenges}
\textbf{Lack of actual user needs driven research}
The empirical studies we survey can be mainly divided into two categories, one is driven by research questions, and the other is driven by user needs. The former usually first raises several research questions which can be composed of independent variables (IVs) and dependent variables (DVs). Generally speaking, IVs are the design method of different XAI interfaces or interaction patterns (see Section 3), and DVs are specific research goals (see Section 4). After identifying the research questions, these studies will design a research prototype interface for a simulated task with some open dataset, and then recruit subjects to participate in the experiment. These research question-driven papers can provide empirical value to the field and can help guide the development of new XAI technologies and the design of XAI interfaces. However, we find that compared with research question-driven studies, user needs-driven studies are rare. User needs-driven studies first obtain the real needs of users through formative studies, interviews, or participatory design methods, to guide the design of the XAI interface. After the design, they evaluate the XAI design via user studies.

\textbf{Lack of user adaptive XAI design}
In XAI design, a recognized concept is that the design of XAI should be user-centric and the difference between users directly affects the effects of XAI. In other words, what effect XAI can achieve is not determined by the XAI designer, but depends on how the user receives and perceives the explanation. Based on this, some works classify the types of users based on their roles in AI applications, or based on their expertise and knowledge background, and summarize different user needs and design goals for different user groups. However, it is not enough to design user group-specific XAI, because even two users who belong to the AI-expert group may be different in many other aspects. Therefore, user-adaptive XAI should be developed.

\subsubsection{Future research opportunities}
\textbf{Complementing research question-driven studies with actual user needs-driven studies.} When conditions permit, researchers can focus more on real users in actual scenarios, and obtain their interpretation needs through user research methods. Although many AI applications have been deployed in people's work and life, they lack effective explanation functions. Researchers can use existing knowledge in the XAI field to solve real-world explanation problems.

\textbf{Building adaptive XAI based on user modeling methods.} Instead of ``one-size-fits-all'' solutions, user-adaptive XAI is expected to adapt the explanations according to users' dynamic information needs, which requires understanding and modeling users in multiple dimensions, including but not limited to the users' ability to perform tasks, AI literacy, domain knowledge, need for cognition, visual literacy, and even the users' inherent impression of AI. Further, the user model needs to be dynamically updated during the interaction.

%% file: sections/designs.tex
\section{Design Space in Current Empirical Study}

In this section, we present an overview of how different aspects and factors are designed or manipulated in existing XAI empirical studies. We believe that this part is critical for new practitioners as this can help them draw a big picture of the current developed design space in this area. The XAI design space is not only for designing different types of explanation methods, but also for other interaction-related factors correlated, such as interface design, and cognitive process design, etc. Note that to make the survey more focused and informative, we focus on empirical studies of XAI, excluding papers that mainly design an explainable system without empirical studies. Based on the surveyed papers, we categorize the design space into Explanation-related design, Model and prediction-related design, and Human-AI collaboration mode-related design.

\subsection{Explanation-related Design.}
Most surveyed papers focus on the study of explanation-related design, varying from the type, and modality to the interactivity and complexity. Figure 1 shows some examples of explanation-based design.

\begin{figure*}[h]
	\centering 
	\includegraphics[width=\linewidth]{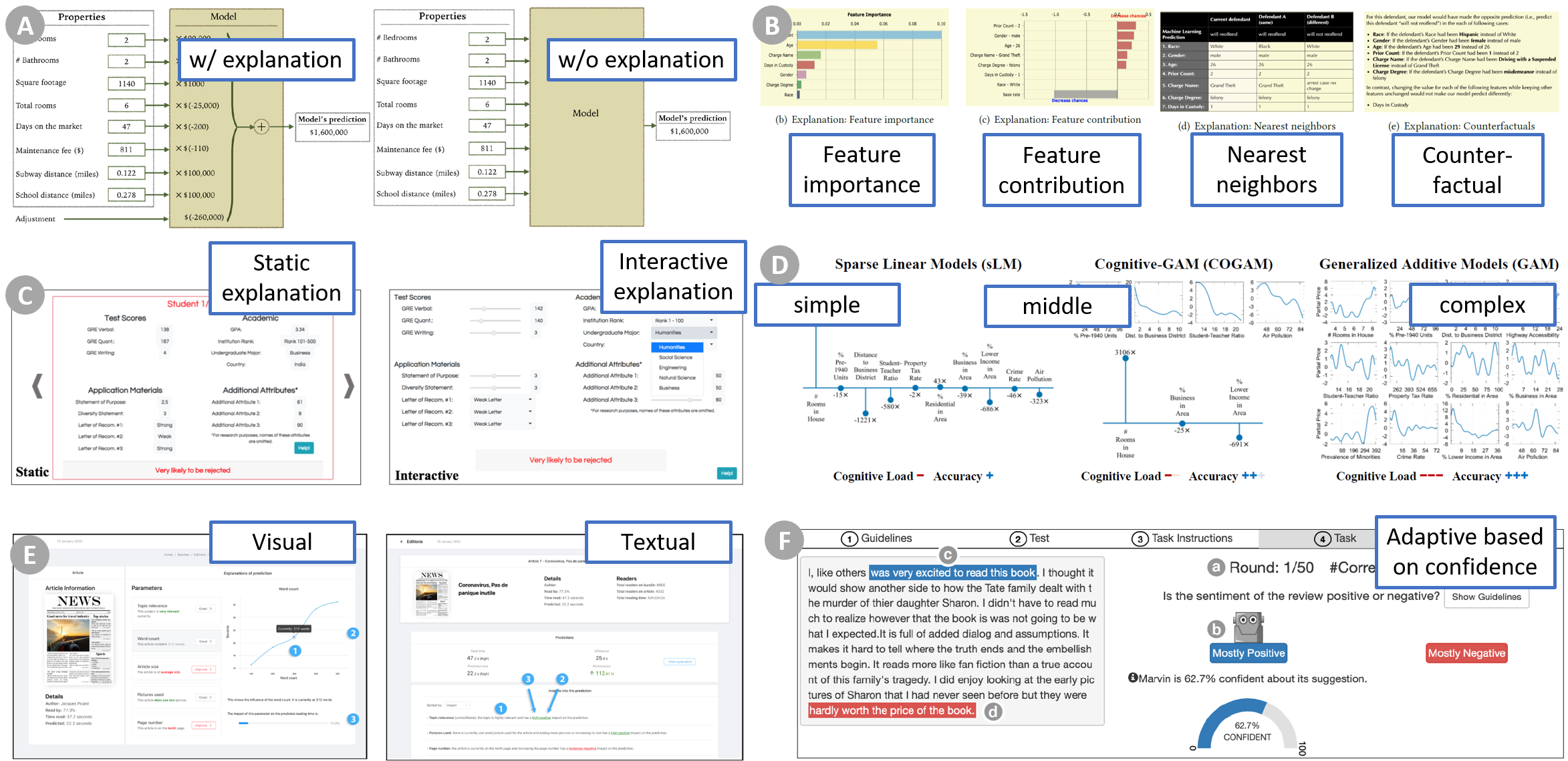}
	\caption{Examples of explanation-related design. (A) w/ and w/o explanation \cite{poursabzi2021manipulating}. (B) Explanation types \cite{wang2021explanations}. (C) Explanation interactivity \cite{cheng2019explaining}. (D) Granularity/complexity of explanation \cite{abdul2020cogam}. (E) Modality of explanation \cite{szymanski2021visual}. (F) Explanation adaptability \cite{bansal2021does}.}
	\label{fig:importance ranking}
\end{figure*}

\subsubsection{D1: W/, W/o, or Random Explanation}
In current XAI empirical studies, a commonly used and intuitive method to investigate the effects of the proposed explanation design is to compare with a ``no-explanation'' condition. Note that the ``no-explanation'' condition in these papers may not be the only baseline, and there might be other baselines to compare with. A great number of works compare one type of explanation with baselines that are without explanations \cite{wang2021explanations, zhang2020effect, fan2021human, zhang2021towards, poursabzi2021manipulating, bansal2021does, cheng2019explaining, robertson2021wait, schaffer2019can}. However, some research has found that even a meaningless explanation or randomly generated explanation can foster users' positive perceptions \cite{ehsan2021explainable, liu2021understanding, lai2019human}. For example, Ehsan et al. \cite{ehsan2021explainable} find that even a meaningless numeric ``explanation'' can lead to over-trust of both AI experts and non-experts. Therefore, to counteract the users' positive attitude toward explanations, some studies use non-informative explanations or random explanations as baselines \cite{ehsan2021explainable, liu2021understanding, lai2019human, ehsan2021explainable}.

\subsubsection{D2: Explanation Type}
As mentioned in Sec. 2, a large number of explanation types has been proposed. To investigate the actual effects of these techniques on users, HCI researchers have designed a lot of explanation types and proposed empirical studies to investigate and compare different types of explanations in targeted tasks \cite{wang2021explanations, zhang2021towards, lai2020chicago, tsai2021exploring, narkar2021model, yang2020visual, dodge2019explaining}. For example, Wang et al. \cite{wang2021explanations} compare four types of common explanations used in literature, including feature importance, feature contribution, nearest neighbors, counterfactuals, and a controlled no-explanation. Recently, Zhang et al. \cite{zhang2021towards} propose a model, RexNet that can generate three types of explanations, Contrastive Saliency, Counterfactual Synthetic, and Contrastive Cues explanations, and compare the effects of different explanation types. Based on a pilot study, Tsai et al. \cite{tsai2021exploring} design three explanation forms for an AI-based COVID-19 symptom checker, including Rationale-based Explanation, Feature-based Explanation, and Example-based Explanation.

% For data scientists, Narkar et al. \cite{narkar2021model} design different types of explanations for an auto-ML model comparison scenario. Yang et al. \cite{yang2020visual} focus on visual explanation and explore the effects of spatial layout and visual representation on users. Dodge et al. \cite{dodge2019explaining} investigate the effects of different explanations on users' fairness judgment of a model. Specifically, they compare Input Influence-based Explanation, Demographic-based Explanation, Sensitivity-based Explanation, and Case-based Explanation.

\subsubsection{D3: Explanation Interactivity}
Interactive explanation is often seen in human-human interaction, which is a social nature of explanation \cite{miller2019explanation}. Some social science works have argued that explanations should be interactive \cite{lombrozo2006structure, miller2019explanation}.

Interactive explanation means that users can interact with the AI's explanation through an interface, such as changing the attribute values of an instance \cite{cheng2019explaining} or creating \emph{What-If} explanations \cite{wang2019designing}, to inspect the updated prediction. The exploration during the interaction can help users understand how the model works. In the context of XAI, Cai et al. \cite{cai2019effects} present a interactive system that allows trial-and-error to explore how an image recognition algorithm works. For a graduate admission prediction task, Cheng et al. \cite{cheng2019explaining} compare the effects of a static explanation and an interactive explanation where users can modify a student's profile to see the changes in prediction. And for human-AI decision making, Liu et al. \cite{liu2021understanding} explore the effects of interactive explanations in three prediction tasks. To help data scientists understand their models, Krause et al. \cite{krause2016interacting} design and implementation of an interactive visual analytics system, Prospector, providing interactive partial dependence diagnostics for users to understand how features affect the prediction.

\subsubsection{D4: Explanation Complexity/Granularity}
The complexity or granularity of an explanation represents the complexity or the amount of detailed information contained in the explanation. In recent studies, researchers begin to pay attention to this dimension. For instance, Lage et al. \cite{lage2019human} conduct a user study to investigate how explanation complexity affects participants’ comprehension and performance. Poursabzi-Sangdeh et al. \cite{poursabzi2021manipulating} manipulate the complexity of a model to alter the interpretability of a model: the number of features used in the model and the transparency of the model. Abdul et al. \cite{abdul2020cogam} proposed COGAM, a model that can adjust the complexity of an explanation based on users' cognitive load. For a sentiment classification task, Springer et al. \cite{springer2019progressive} compare different explanations with different levels of complexity. Mishra et al. \cite{mishra2021crowdsourcing} explore granularity (of data features) and context (of data instances) as dimensions, and investigate the effects of granularity, context, simplification affect user understanding and confidence in ML models.

\subsubsection{D5: Explanation Modality}

The modality in which explanations are presented is an important characterization affecting how users will perceive it \cite{mohseni2021multidisciplinary, szymanski2021visual}. Explanations could be presented in multi-modality, such as textual \cite{tsai2021exploring, wilkinson2021or, rastogi2020deciding}, verbal \cite{hohman2019telegam}, visual in the form of graphs (e.g. saliency maps \cite{adadi2018peeking}, scatterplot diagrams \cite{krause2016interacting}, bar chart \cite{zhang2020effect}, attention map \cite{lai2020chicago}, line charts \cite{wang2019designing}), etc. Some researchers have begun to be interested in explained modalities \cite{park2018multimodal, contreras2018artificial}. For example, in a recommender system, Kouki et al. \cite{kouki2019personalized} find that textual explanations are more persuasive than visual explanations. Hohman et al. \cite{hohman2019telegam} combine visual explanations with textual explanations to vary the complexity. Szymanski et al. \cite{szymanski2021visual} propose three modalities of explanation, textual explanation, visual explanation, and hybrid explanation. Through a user study on video game players, Robertson et al. \cite{robertson2021wait} find that when users' attention resources are limited, presenting explanation interventions via different modalities, like audio, interactivity, and text, can aid real-time comprehension. Due to the limited number of research on the modality of explanation, more efforts are needed to explore how and when to combine multi-modal explanations for a unified goal.

\subsubsection{D6: Explanation Adaptability}
Besides the traditional pre-defined explanations, from the surveyed papers published in recent two years, we also find several works proposing adaptive explanations, which show a promising direction for better dynamic explanations to cater to users' various cognitive abilities and needs. For example, Bansal et al. \cite{bansal2021does} introduce an adaptive explanation. It tries to reduce human trust when the AI has low confidence: it only explains the predicted class (Top-1-explanation) when the AI is confident, but also explains the alternative (Top-2-explanation) otherwise. Rastogi et al. \cite{rastogi2020deciding} propose a confidence-based adaptive time allocation strategy to allocate a different amount of time for users to perceive and understand an explanation. Abdul et al. \cite{abdul2020cogam} propose COGAM, an adaptive explanation complexity adjustment method by calibrating visual cognitive chunks with users' potential cognitive load, and the COGAM can achieve the trade-off between cognitive load and accuracy.

% \subsubsection{D7: Explanation for One or More Classes}
% In general, taking the classification task as an example, the AI model will generate the explanation for the predicted class by default. However, the AI can also generate explanations for other alternative classes. Some works have shown that only offering explanations for the single predicted class will increase users' reliance on recommendations even when the AI was incorrect \cite{bansal2021does}. It will lead to inappropriate trust on systems \cite{kaur2020interpreting, mitchell2019model}, i.e., explanations can lead humans to either follow incorrect AI suggestions or ignore the correct ones \cite{bussone2015role}. As a first attempt to tackle the problem of blind reliance on AI, Bansal et al. \cite{bansal2021does} introduce an adaptive explanation. It tries to reduce human trust when the AI has low confidence: it only explains the predicted class (Top-1-explanation) when the AI is confident, but also explains the alternative (Top-2-explanation) otherwise.

\subsection{Model and Prediction-related Design.}
In addition to the design directly related to explanations, there is some information that can also play a role in helping people understand AI, including information related to models and predictions. Figure 2 shows the examples of model and prediction-related design.

\begin{figure*}[h]
	\centering 
	\includegraphics[width=0.8\linewidth]{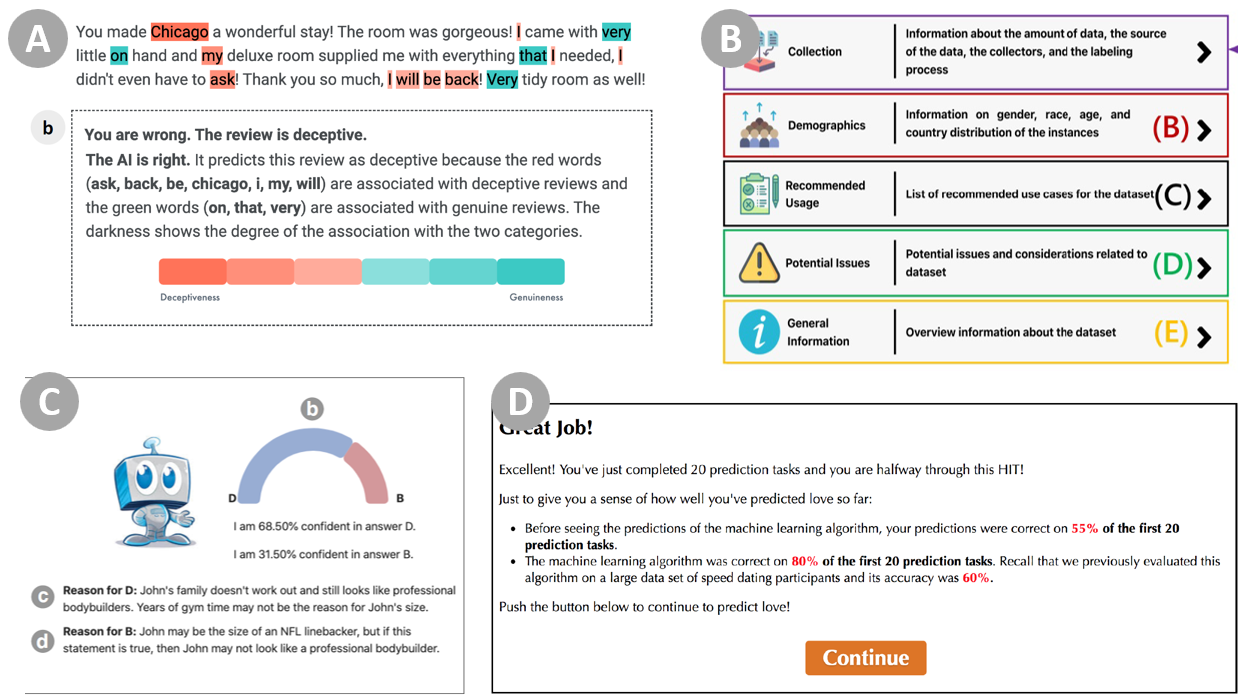}
	\caption{Model and prediction-related design. (A) W/ and W/o Tutorial/Training \cite{lai2020chicago}. (B) Explanation of Training Data \cite{anik2021data}. (C) W/ and W/o Confidence/Uncertainty \cite{bansal2021does}. (D) Model performance \cite{yin2019understanding}.}
	\label{fig:importance ranking}
\end{figure*}

\subsubsection{D7: W/ and W/o Tutorial/Training}
Some studies have found that only presenting global or local explanations during the task process is not effective enough for users to fully understand the AI \cite{lai2020chicago, arnold2019factsheets, mitchell2019model}. They pay attention to the tutorial process before the task. For example, instead of directly providing real-time assistance to people, Lai et al. \cite{lai2020chicago} focus on the training stage, and propose tutorials to help users understand the patterns embedded in a model and the nature of a task. Particularly, they propose two kinds of tutorials, guideline-based and example-driven tutorials.

\subsubsection{D8: Explanation of Training Data}

Some research finds that providing the information of training data can facilitate users' understanding of the model, such as the input features used or data distribution \cite{dodge2019explaining, harrison2020empirical, kulesza2013too, poursabzi2021manipulating, zhang2020effect}. For example, in an income prediction task, Zhang et al. \cite{zhang2020effect} investigate the effect of humans are made aware of whether or not the model considers ``marital status'' as a feature. Liu et al. \cite{liu2021understanding} investigate the effects of telling users the data distribution on human-AI decision making. To promote transparency, Anik et al. \cite{anik2021data} propose data-centric explanations to explain a series of information of training data, including how the data was collected, demographics, recommended usage, potential issues, etc.

\subsubsection{D9: W/ and W/o Confidence/Uncertainty}
Controlling whether showing a model's confidence or uncertainty is one of the most common designs in human-AI collaborative decision-making studies. For example, Zhang et al. \cite{zhang2020effect} investigate the effects of confidence scores on users' trust, and accuracy of AI-assisted predictions. Rastogi et al. \cite{rastogi2020deciding} propose a confidence-based time allocation strategy for AI-assisted decision-making, and find that when the AI model has low confidence and is incorrect, their proposed confidence-based time allocation strategy can effectively mitigate users' cognitive biases and improve the collaborative performance. Bansal et al. \cite{bansal2021does} compare different explanations with a confidence-based condition and find that confidence scores can potentially help people form a good mental model of AI’s error boundaries.

\subsubsection{D10: Model performance}
Another important piece of information is the model's performance in the training dataset, which is shown to be able to affect users' trust. In classification tasks, model performance is mainly presented in the form of accuracy \cite{harrison2020empirical, lai2020chicago, lai2019human, yang2020visual, yin2019understanding}. There have been some studies exploring how showing model accuracy to users affects their perception of and task performance. For example, Lai et al. \cite{lai2020chicago} investigate whether the model's accuracy improves users' performance in decision-making tasks. And, Yin et al. \cite{yin2019understanding} study the effect of accuracy on human trust in ML models. Model performance can also be measured by precision and recall. For example, in a recidivism prediction task, Harrison et al. \cite{harrison2020empirical} show that presenting false positive rates can help people judge the fairness of the model. Considering that model performance estimated in the training data can be inconsistent with the real performance, Yin et al. \cite{yin2019understanding} investigate the effect of communicated accuracy and experienced accuracy on users.

\subsection{Human-AI Collaboration Mode-related Design.}
% Or human cognition-related design.
Figure 3 shows the examples of human-AI collaboration mode-related design.

\begin{figure*}[h]
	\centering 
	\includegraphics[width=0.8\linewidth]{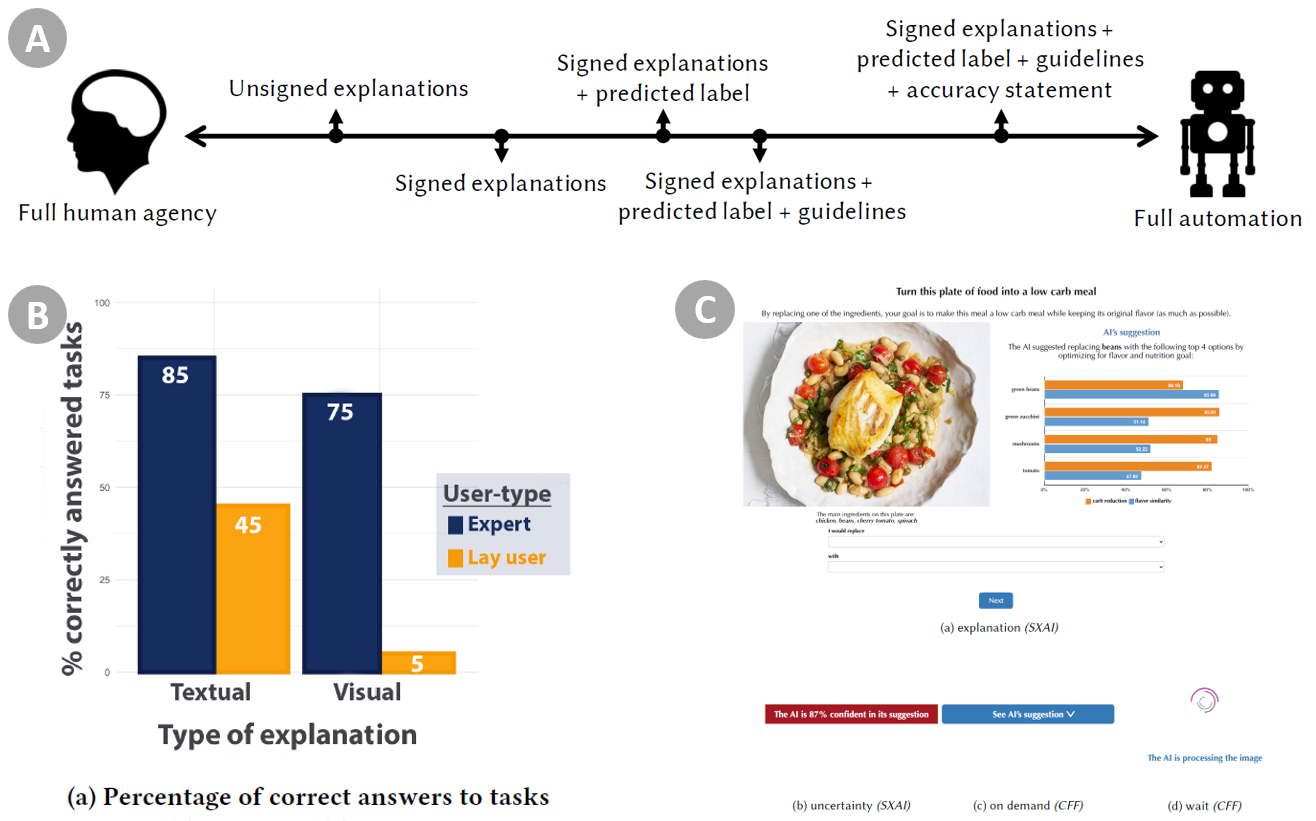}
	\caption{Human-AI collaboration mode-related design. (A) AI Agency (Degree of Explanation) \cite{lai2019human}. (B) Expertise of Explanation Users \cite{szymanski2021visual}. (C) Cognitive Bias Mitigating \cite{buccinca2021trust}.}
	\label{fig:importance ranking}
\end{figure*}

\subsubsection{D11: AI Agency (Degree of Explanation)}
AI can have different levels of agency in the human-AI interaction process. In the context of XAI, the agency can be determined by the degree of explanation. For example, a model only with an explanation is a low agency, while a model with prediction, confidence, and explanation is high agency. Researchers have compared different levels of machine agency. For example, Buçinca et al. \cite{buccinca2021trust} examine the effects of on-demand explanation that is receiving predictions only on demand. For a deception detection task, Lai et al. \cite{lai2019human} compare different AI agencies, including 1) Control-human only, 2) Feature-based explanations, 3) Example-based explanations, 4) Predicted label without accuracy, 5) Predicted label with accuracy, 6) Combinations. Levy et al. \cite{levy2021assessing} compare two clinical note annotation systems with different levels of AI agency, one only suggests annotation labels after users choose text spans to be labeled, and another performs both span and label suggestions.

\subsubsection{D12: Expertise of Explanation Users}

It has been well recognized in the XAI area that user expertise is a key aspect to be considered when designing an explanatory interface \cite{arrieta2020explainable, burnett2020explaining, sperrle2020should}. It is an open challenge in XAI to tailor the explanations to the expertise of the end-user \cite{gunning2019xai}.

In the empirical studies that focus on user expertise as an independent variable, Szymanski et al. \cite{szymanski2021visual} investigate how different levels of expertise influence the understanding of explanations. Wang et al. \cite{wang2021explanations} investigate the effects of different explanations on two tasks where participants have more expertise in one task but not in the other. Anik et al. \cite{anik2021data} test their proposed explanation on users with different expertise, i.e., expert, intermediate, beginner. In a graduate admission prediction task, Cheng et al. \cite{cheng2019explaining} explore how the effectiveness of the explanation interface can be influenced by users’ personal characteristics (i.e. education level and technical literacy). Besides, Schaffer et al. \cite{schaffer2019can} explore the difference between participants' measured expertise and their self-reported familiarity in an AI-assisted binary decision-making game. In addition, Ehsan et al. \cite{ehsan2021explainable} conduct a mixed-methods study to investigate how users with different AI backgrounds perceive different types of AI explanations.

\subsubsection{D13: Cognitive Bias Mitigating}

How humans perceive the explanation is at the core of whether and how an explanation works for them \cite{rastogi2020deciding}. However, human cognition is inclined to natural flaws, like cognitive biases \cite{zhang2021towards}. Recently, a number of empirical studies have focused on designing cognitive de-bias methods to better communicate the explanation to the user \cite{rastogi2020deciding, buccinca2021trust}. Evidence shows that providing interventions to change users' cognitive processes can significantly affect how they process the explanation information to form mental models of AI and make decisions \cite{lai2021towards}.

One method to mitigate the cognitive bias is designing a more appropriate workflow. The typical paradigm of AI-assisted interaction is to have the AI provide its suggestion, then the user can choose to take the AI's advice or follow her own decision. Some studies explore the effects of having users make their own predictions before being shown the model output \cite{buccinca2021trust, grgic2019human, lu2021human, poursabzi2021manipulating, wang2021explanations, yin2019understanding, zhang2020effect}. Such designs are verified to be able to force people to think more analytically rather than merely relying on the model's predictions. Existing work also explores the impact of different workflow designs on users’ mental models, such as including a training phase before the task \cite{chandrasekaran2018explanations, lai2020chicago, poursabzi2021manipulating, zhang2020effect}, showing users their own or models’ decision correctness \cite{bansal2019updates, bansal2021does, chandrasekaran2018explanations, grgic2019human, yang2020visual, yu2019trust}. Focusing on the anchoring bias, Nourani et al. \cite{nourani2021anchoring} investigate how the order of observing system weaknesses and strengths can affect the user’s mental model, task performance, and reliance.

Another method to mitigate the cognitive bias is modifying the system response time \cite{buccinca2021trust, park2019slow}. For example, Buçinca et al. \cite{buccinca2021trust} propose and explore the effects of three kinds of cognitive forcing functions. One of them is slowing down the process that the AI offers its prediction. Rastogi et al. \cite{rastogi2020deciding} propose an adaptive time-allocation method and find that allocating more time can alleviate anchoring bias, especially in the condition that AI makes a mistake.

% \subsection{Explanation-related Design.}
% \subsubsection{D1: W/, W/o, or Random Explanation}
% \subsubsection{D2: Explanation Type}
% \subsubsection{D3: Explanation Interactivity}
% \subsubsection{D4: Explanation Complexity/Granularity}
% \subsubsection{D14: Explanation Modality}
% \subsubsection{D15: Explanation Adaptability}

% \subsection{Model and Prediction-related Design.}
% \subsubsection{D5: W/ and W/o Tutorial/Training}
% \subsubsection{D9: Explanation of Training Data}
% \subsubsection{D10: W/ and W/o Confidence/Uncertainty}
% \subsubsection{D16: Model performance}

% \subsection{Human-AI Collaboration Mode-related Design.}
% \subsubsection{D6: AI Agency (Degree of Explanation)}
% \subsubsection{D8: Expertise of Explanation Users}
% \subsubsection{D12: Cognitive Bias Mitigating}

% \subsection{Evaluation related to users' perceptions of AI}
% \subsubsection{}
% \subsubsection{G3: Trust/Reliance}
% \subsubsection{G6: Trust Calibration}
% \subsubsection{G9: Fairness Judgement}

% \subsection{Evaluation related to users' interaction experience.}
% \subsubsection{G4: Satisfaction/Perception of the system.}
% \subsubsection{G8: Cognitive Load/Mental Demand}
% \subsubsection{G11: Engagement}

% \subsection{Evaluation related to the task}
% \subsubsection{G2: Task Performance}
% \subsubsection{G7: Time Spent/Task Efficiency}

% \begin{landscape}
\renewcommand{\arraystretch}{1.2}
\begin{table}[tp]  
  \centering  
  \fontsize{8}{8}\selectfont

\begin{threeparttable}  

\caption{Summary of a subset of the surveyed literature organized by the two dimensions, design Space (independent variable) and evaluation goal (dependent variable)}\label{table:posttask-questionnaire-student} 
% \begin{tabular}{l|l|l|l|l|l|l|l|l|l|l|l|l|l|l|l|l|l|l|l|l|l|l|l}
\begin{tabular}{p{3cm}|p{0.2cm}p{0.2cm}p{0.2cm}p{0.2cm}p{0.2cm}p{0.2cm}|p{0.2cm}p{0.2cm}p{0.2cm}p{0.2cm}|p{0.2cm}p{0.2cm}p{0.2cm}|p{0.2cm}p{0.2cm}p{0.2cm}p{0.2cm}|p{0.2cm}p{0.2cm}p{0.2cm}|p{0.2cm}p{0.2cm}}
%\begin{tabular}{cccccc}

\toprule
&\multicolumn{13}{c|}{\normalsize \textbf{Design Space}}&\multicolumn{9}{c}{\normalsize \textbf{Evaluation Goal}}\\
&\multicolumn{6}{p{2.5cm}|}{\textbf{Explanation-related Design}}&\multicolumn{4}{p{1.8cm}|}{\textbf{Model and Prediction-related Design}}&\multicolumn{3}{p{1.3cm}|}{\textbf{Collaboration Mode-related Design}}&\multicolumn{4}{p{1.8cm}|}{\textbf{Evaluation related to users' perceptions of AI}}&\multicolumn{3}{p{1.3cm}|}{\textbf{Evaluation related to users' interaction experience}}&\multicolumn{2}{p{1cm}}{\textbf{Evaluation related to the task}}\\

\normalsize \textbf{Work}&\rotatebox{90}{\textbf{D1: W/, W/o, or Random Explanation}}&\rotatebox{90}{\textbf{D2: Explanation Type}}&\rotatebox{90}{\textbf{D3: Explanation Interactivity}}&\rotatebox{90}{\textbf{D4: Explanation Complexity/Granularity}}&\rotatebox{90}{\textbf{D5: Explanation Modality}}&\rotatebox{90}{\textbf{D6: Explanation Adaptability}}&\rotatebox{90}{\textbf{D7: W/ and W/o Tutorial/Training}}&\rotatebox{90}{\textbf{D8: Explanation of Training Data}}&\rotatebox{90}{\textbf{D9: W/ and W/o Confidence/Uncertainty}}&\rotatebox{90}{\textbf{D10: Model performance}}&\rotatebox{90}{\textbf{D11: AI Agency (Degree of Explanation)}}&\rotatebox{90}{\textbf{D12: Expertise of Explanation Users}}&\rotatebox{90}{\textbf{D13: Cognitive Bias Mitigating}}&\rotatebox{90}{\textbf{G1: Understanding/Mental Model}}&\rotatebox{90}{\textbf{G2: Trust/Reliance}}&\rotatebox{90}{\textbf{G3: Trust Calibration}}&\rotatebox{90}{\textbf{G4: Fairness Judgment}}&\rotatebox{90}{\textbf{G5: Satisfaction/Perception of the system}}&\rotatebox{90}{\textbf{G6: Cognitive Load/Mental Demand}}&\rotatebox{90}{\textbf{G7: Engagement}}&\rotatebox{90}{\textbf{G8: Task Performance}}&\rotatebox{90}{\textbf{G9: Time Spent/Task Efficiency}}\\ \hline
% xxxx&&&&&&&&&&&&&&&&\\
\rowcolor{LightCyan}
Wilkinson et al. 2021 \cite{wilkinson2021or}&&\checkmark&&&&&&&&&&&&\checkmark&\checkmark&&&&&&&\\

Szymanski et al. 2021 \cite{szymanski2021visual}&&&&&\checkmark&&&&&&&\checkmark&&\checkmark&&&&\checkmark&&&&\\
\rowcolor{LightCyan}
Chromik et al. 2021 \cite{chromik2021think}&&&&&&&\checkmark&&&&&&&\checkmark&&&&&&&&\\
Wang et al. 2021 \cite{wang2021explanations}&\checkmark&\checkmark&&&&&&&&&&\checkmark&&\checkmark&&\checkmark&&&&&&\\
\rowcolor{LightCyan}
Buçinca et al. 2020 \cite{buccinca2020proxy}&\checkmark&\checkmark&&&&&&&&&&&&\checkmark&\checkmark&&&\checkmark&&&\checkmark&\\
Zhang et al. 2020 \cite{zhang2020effect}&\checkmark&&&&&&&&\checkmark&&&&&&&\checkmark&&&&&\checkmark&\\
\rowcolor{LightCyan}
Lai et al. 2019 \cite{lai2019human}&&&&&&&&&&&\checkmark&&&&\checkmark&&&&&&\checkmark&\\
Fan et al. 2022 \cite{fan2021human}&\checkmark&&&&&&&&&&&&\checkmark&\checkmark&\checkmark&&&\checkmark&&\checkmark&\checkmark&\\
\rowcolor{LightCyan}
Rastogi et al. 2022 \cite{rastogi2020deciding}&&&&&&\checkmark&&&\checkmark&&&&\checkmark&&&&&&&&\checkmark&\\
Buçinca et al. 2021 \cite{buccinca2021trust}&&&&&&&&&\checkmark&&&&\checkmark&&&\checkmark&&\checkmark&\checkmark&&\checkmark&\\
\rowcolor{LightCyan}
Zhang et al. 2022 \cite{zhang2021towards}&\checkmark&\checkmark&&&&&&&&&&&&\checkmark&&&&\checkmark&&&\checkmark&\checkmark\\
Sangdeh et al. 2021 \cite{poursabzi2021manipulating}&\checkmark&&&\checkmark&&&&&&&&&&\checkmark&&\checkmark&&&&&&\\
\rowcolor{LightCyan}
Bansal et al. 2021 \cite{bansal2021does}&\checkmark&&&&&\checkmark&&&\checkmark&&&&&&&&&\checkmark&&&\checkmark&\\
Tsai et al. 2021 \cite{tsai2021exploring}&&\checkmark&&&&&&&&&&&&\checkmark&\checkmark&&&\checkmark&\checkmark&&&\\
\rowcolor{LightCyan}
Anik et al. 2021 \cite{anik2021data}&&&&&&&&\checkmark&&&&\checkmark&&&\checkmark&&\checkmark&\checkmark&&&&\\
Lai et al. 2020 \cite{lai2020chicago}&&\checkmark&&&&&\checkmark&&&\checkmark&\checkmark&&&&&&&\checkmark&&&\checkmark&\\
\rowcolor{LightCyan}
Abdul et al. 2020 \cite{abdul2020cogam}&&&&\checkmark&&\checkmark&&&&&&&&\checkmark&&&&&\checkmark&&&\\
Cheng et al. 2019 \cite{cheng2019explaining}&\checkmark&&\checkmark&&&&&&&&&\checkmark&&\checkmark&\checkmark&&&&&&&\checkmark\\
\rowcolor{LightCyan}
Robertson et al. 2021 \cite{robertson2021wait}&\checkmark&\checkmark&\checkmark&&\checkmark&&&&&&&&&\checkmark&&&&&\checkmark&&&\\
Narkar et al. 2021 \cite{narkar2021model}&&\checkmark&&&&&&&&&&&&&&&&\checkmark&&&&\\
\rowcolor{LightCyan}
Nourani et al. 2021 \cite{nourani2021anchoring}&&&&&&&&&&&&&\checkmark&\checkmark&\checkmark&&&&&&\checkmark&\\
Yang et al. 2020 \cite{yang2020visual}&&\checkmark&&&&&&&&&&&&&&\checkmark&&&&&&\\
\rowcolor{LightCyan}
Springer et al. 2019 \cite{springer2019progressive}&&&&\checkmark&&&&&&&&&&\checkmark&\checkmark&&&&&&&\\
Schaffer et al. 2019 \cite{schaffer2019can}&\checkmark&&&&&&&&&&&\checkmark&&&\checkmark&&&&&&&\\
\rowcolor{LightCyan}
Dodge et al. 2019 \cite{dodge2019explaining}&&\checkmark&&&&&&&&&&&&&&&\checkmark&&&&&\\
Liu et al. 2021 \cite{liu2021understanding}&&&\checkmark&&&&&\checkmark&&&&&&&&&&\checkmark&&&\checkmark&\\
\rowcolor{LightCyan}
Mishra et al. 2021 \cite{mishra2021crowdsourcing}&&&&\checkmark&&&&&&&&&&\checkmark&&&&&&&&\\
Ehsan et al. 2021 \cite{ehsan2021explainable}&&&&&&&&&&&&\checkmark&&\checkmark&\checkmark&&&\checkmark&&&&\\
\rowcolor{LightCyan}
Yin et al. 2019 \cite{yin2019understanding}&&&&&&&&&&\checkmark&&&&&\checkmark&&&\checkmark&&&&\\
% xxxx&&&&&&&&&&&&&&&&\\

\bottomrule
\end{tabular}
\end{threeparttable}  

\end{table}

% \end{landscape}

\subsection{Summary and Thinking}
Existing human-centered XAI research has explored a wide design space, which fall into explanation-related design, mode and prediction-related design, human-AI collaboration model-related design. An overview of the design space can help practitioners make better design choices.

\subsubsection{Challenges}
\textbf{Lack of XAI design for the actual scene}
Currently, most XAI empirical studies simulate real task scenarios by designing an experimental interface. However, there are several differences between this experimental process and the real task scene. The first difference is in the task context and the process. In the empirical study, a simplified task process is used. After being told the experimental task, the user will directly enter the decision-making trials. However, in the actual task, other steps that seem not directly related to the decision-making are likely to affect the user's perception of XAI. The second difference is between experimental data and real data. Due to some limitations, most experiments are now based on public data sets. However, the composition of these public data sets could be different from real-world data. The third difference is the end user. Many studies now choose to use crowdsourcing experimental methods to obtain sufficient data. However, it can lead to some problems, such as fidelity, especially when the task setting requires a certain professional background. With these three differences, XAI design can be evaluated in an inappropriate manner and the findings may not be able to generalize to real-world tasks.

\textbf{Lack of modeling of human mental model}
There is now a substantial body of work that emphasizes calibrating users' mental models \cite{buccinca2021trust, bansal2021does}. However, to calibrate the user's mental model, an essential premise is to know clearly what the user's current mental model is. Taking user trust calibration as an example, many studies have realized that people should increase trust when the AI makes correct predictions, and decrease trust when the AI makes mistakes, but they have not studied the trust level of users at the moment. If the user's trust in AI is already very low, there is no need to try to reduce the user's trust by providing additional information such as an explanation or low confidence.

\textbf{A limited exploration and exploitation of design space}
On the one hand, the design space has not been fully exploited. A lot of design elements in the existing design space have not been fully investigated, such as modality and interactivity. In addition, there is relatively little knowledge of how the combination of different dimensions can affect users and how different dimensions interplay with each other. On the other hand, the existing design space is far from being fully explored. In each dimension, there can be plenty of new design elements. Also, there are many other possible design dimensions worth exploring.

\textbf{Limited human-AI cooperation mode in XAI study}
Most XAI research now focuses on decision-making or recommendation scenarios. We realize that there are many other cooperation modes, such as multi-round collaboration mode and long-term collaboration mode, which require the design of XAI to consider changes in people's mental models over time, as well as the long-term impact of XAI on people.

\textbf{Lack of exploration on the role of XAI in other human-centered aspects}
The explainability of AI can help address other aspects of human-centered problems. For example, XAI can facilitate assessments of AI security, audit fairness, design more private AI, and explore accountability issues. However, we find that existing XAI research pays less attention to these aspects.

\subsubsection{Future research opportunities}

\textbf{Exploring human-centered XAI design in more actual scenario.} Researchers can design XAI systems/interfaces for real-world applications. In an alternative way, researchers can try to recruit actual stakeholders for their research prototype. This can not only enrich the current XAI research space but also bring unique insights and practical implications for a specific type of user.

\textbf{Design methods to model human mental model.} Since the human mental model can determine the effects of explanations, modeling users' mental model during the interaction process is promising. At the same time, we also recognize its difficulty, because it is difficult to quantify the user's mental model. Explicit modeling methods, such as asking the user to answer a few questions to measure the user's mental model during the execution of the task, will seriously damage the user's experience. While implicit methods (such as modeling the user through the user's interaction data in the interface, the time of hesitation, or the user's expression, etc.) are likely to face the problem of inaccuracy. We believe that this is a very worthy research direction in the future.

\textbf{Exploring new XAI design.} The purpose of summarizing the existing design space is not to limit the ideas of designers, but to let designers realize the insufficiency and blankness of the current design space. In the future, researchers should explore new design dimensions or explore new design elements or combinations between different design elements. And empirical research can serve as the ground for exploration.

\textbf{Investigating the effects of XAI beyond one-round human-AI interaction.} Most of the current XAI research uses a one-round of human-AI interaction mode, or the same sub-task is repeated many times. We find that there are still research gaps in the multi-round interaction. Researchers can explore this area in the future. There are specific research questions to explore, such as for a multi-round decision-making task, how the explanation provided by the AI should change as the interaction progresses to keep calibrate users' trust, and how the user's decision changes after seeing updated AI explanations in a new round.

\textbf{Investigating the effects of XAI beyond one-on-one human-AI interaction.} The current experimental scenarios are almost all one person interacting with one AI, and little work has been done to explore how multi-person and multi-AI teams interact internally. Taking decision-making tasks as an example, in a scenario of two people and two AIs, there will be many different voting situations, so how should the final decision be determined? How do team members communicate? What are the roles of humans and AI members in team decision-making? These issues are worth exploring in the future.

%% file: sections/evaluations.tex
\section{Evaluation of XAI Design}
Evaluation for XAI systems is a critical step in the XAI empirical studies. Explanations or explanatory interfaces are designed to achieve different interpretability goals (based on broad user needs or research questions), and hence different measures are needed for the intended purpose \cite{mohseni2021multidisciplinary}. In this section, we focus on evaluation metrics. Reviewing the surveyed papers, we summarize a list of evaluation measures (\emph{how to evaluate}) associated with their evaluation goals (\emph{what to evaluate}), as shown in Tables 3, 4, 5.

At a high level, we group the evaluation metrics into three categories: (1) evaluation related to users' perceptions of AI, (2) evaluation related to users' interaction experience, and (3) evaluation related to the task. Under each, we categorize metrics based on evaluation goals and further classify them based on subjectivity. We note that the qualitative methods (such as interviews and think-aloud) can be diverse in each work and few works provide the question list in interviews in their papers. Thus, here we only review quantitative methods used in surveyed papers.

\subsection{Evaluation Related to Users' Perceptions of AI}
Since one of the most direct design goals of XAI is to help users understand the AI, a lot of evaluation is situated on this goal, including users' understanding/mental model of AI, trust/reliance in AI, trust calibration, fairness judgment, and others. Table 3 summarizes these measures.

\subsubsection{Understanding/Mental Model.}

Users' understanding of the AI is the most investigated evaluation goal in the surveyed papers. Subjective metrics includes users' \emph{self-report of understanding of the AI} \cite{anik2021data, buccinca2020proxy, cai2019effects, cheng2019explaining, lucic2020does, wang2021explanations, yang2020visual}, \emph{confidence in understanding} \cite{kulesza2012tell}, \emph{ease of understanding} \cite{poursabzi2021manipulating}, \emph{confidence in simulation} \cite{alqaraawi2020evaluating, nguyen2018comparing}, \emph{perceived intuitiveness} \cite{szymanski2021visual} or \emph{perceived transparency} \cite{rader2018explanations, tsai2021exploring} of the AI system. Generally, in our surveyed studies, researchers often use subjective questionnaire (usually 5 or 7-point Likert scale) to measure users' understanding or users' perceived transparency of the AI \cite{wilkinson2021or, szymanski2021visual, fan2021human, tsai2021exploring, mishra2021crowdsourcing, ehsan2021explainable, buccinca2020proxy}.

Objective metrics is often used to test how well users understand \emph{how the system works} and \emph{what the output will be} compared to actual ground-truth. The most commonly used metric is forward simulation \cite{abdul2020cogam, buccinca2020proxy, chromik2021think, hase2020evaluating, nourani2021anchoring, poursabzi2021manipulating, wang2021explanations}, which asks participants to guess a model’s predictions on unseen instances. In a slightly different manner, counterfactual simulation is also used in some studies \cite{hase2020evaluating, wang2021explanations}. Apart from simulation, other objective metrics has been utilized, such as the correctness of people’s assessment of model performance \cite{nourani2021anchoring, springer2019progressive}, identification of important features \cite{cheng2019explaining, szymanski2021visual}, detection of errors \cite{wang2021explanations, poursabzi2021manipulating}, or description of model behaviors \cite{chromik2021think}. These measures are often conducted by quizzes \cite{cheng2019explaining, gero2020mental, robertson2021wait, wang2021explanations}.
For example, Robertson et al. \cite{robertson2021wait} measure users' understanding by asking participants to answer multiple-choice recall questions.

Some studies adopt both subjective measures and objective measures \cite{cheng2019explaining, chromik2021think, wang2021explanations}. For example, in a comparison study of different types of explanation, Wang et al. \cite{wang2021explanations} measure users' both subjective understanding and objective understanding. For the subjective measures, they ask participants to self-report their understanding on a Likert scale. For the objective measures, they design five kinds of quiz questions. Note that objective and subjective understanding does not always align due to some reasons, such as ``illusory confidence'' that humans believe they understood the model more than they actually did \cite{chromik2021think, cheng2019explaining}.

%wang2021explanations
% Typical methods include ask people to rank the input data features based on their influence to overall predictions \cite{collaris2018instance, hohman2019gamut}[18, 33], to indicate the direction of change in the model’s prediction when a feature’s value is altered \cite{cheng2019explaining, collaris2018instance, goldstein2015peeking}[16, 18, 29], to simulate the model’s predictions \cite{cheng2019explaining, doshi2017towards, lage2019evaluation, lipton2018mythos, poursabzi2021manipulating} [16, 24, 40, 47, 59], to answer “what-if” questions about the model behavior \cite{binns2018s, cheng2019explaining, doshi2017towards, hohman2019gamut, miller2019explanation} [7, 16, 24, 33, 54], and to detect mistakes of the model and debug the model \cite{poursabzi2021manipulating, ribeiro2016should} [59, 60].

\subsubsection{Trust/Reliance}
Trust in AI is an essential research topic. There are some types of subjective measures, such as direct self-reported trust \cite{abdul2020cogam, buccinca2020proxy, cheng2019explaining, chromik2021think, poursabzi2021manipulating, springer2019progressive, tsai2021exploring}, self-reported agreement or reliance \cite{chandrasekaran2018explanations}, acceptance or confidence in the model \cite{chromik2021think, szymanski2021visual, wang2021explanations}, or perceived accuracy of the AI \cite{kocielnik2019will, smith2020no, springer2019progressive}. Most works in Table 3 measure users' trust subjectively. For example, Wilkinson et al. \cite{wilkinson2021or} (in a movie recommendation task), Buçinca et al. \cite{buccinca2020proxy} (in a food percent fat content of nutrients recognition task), Lai et al. \cite{lai2019human, lai2020chicago} (in a deceptive review detection task), Cheng et al. \cite{cheng2019explaining} (in a graduate admission prediction task), measure users' trust by subjective questionnaire.

Objective metrics focus more on reliance (i.e., whether users take the model's advice or how much users' decisions are influenced by the AI’s suggestions), such as acceptance of model suggestions \cite{bansal2021does, lai2020chicago, lai2019human, liu2021understanding, lu2021human, wang2021explanations, yin2019understanding, zhang2020effect}, likelihood to switch \cite{lu2021human, yin2019understanding, zhang2020effect}, weight of model advice \cite{poursabzi2021manipulating}, as well as deviation from the model’s suggestions \cite{poursabzi2021manipulating}. For example, in an income prediction task, Zhang et al. \cite{zhang2020effect} use two indicators to measure users' trust, \emph{switch percentage} which means how many users' decisions switch to AI's prediction, and \emph{agreement percentage} which means how many users' decisions are aligned with AI's prediction.

It is worth noting that trust and reliance are not the same. Trust is more like an attitude and reliance is more close to a behavior. Apart from trust, reliance can be influenced by other factors, such as efforts to engage, time constraints, perceived risk, workload, and self-confidence\cite{lee2004trust, lai2021towards}. There are also some surveys that conclude how to measure users' trust in empirical studies \cite{lai2021towards, mohseni2021multidisciplinary, vereschak2021evaluate}.

% For example, drawing from empirical practices in social and cognitive studies on human-human trust, Vereschak et al. \cite{vereschak2021evaluate} provide practical guidelines to improve the methodology of studying Human-AI trust in decision-making contexts.

\subsubsection{Trust Calibration}
% this section is ok.
Here, we deliberately highlight this goal separately. Although \emph{trust calibration} can be seen as a subset of \emph{trust}, most studies in \emph{trust} aim to increase trust in users by providing an explanation. However, in recent days, more and more research's focus has changed from increasing users' trust to maintaining an appropriate trust in the AI. Because to achieve a complementary human-AI team performance \cite{bansal2019beyond, bansal2021does, rastogi2020deciding, buccinca2020proxy}, an essential step is to guide people to trust or to be cautious in different situations with the help of explanations (trust calibration \cite{yang2020visual, zhang2020effect}).

To evaluate the effects of explanation design on trust calibration, researchers usually measure it in two cases, one is whether people make decisions that are consistent with AI when AI makes correct decisions, and the other is whether people doubt AI when AI makes wrong decisions. Thus, existing empirical studies propose three objective measures, (1) the over-reliance \cite{buccinca2021trust, wang2021explanations, yang2020visual} which means the human follows the AI's prediction when the AI is wrong; (2) under-reliance \cite{wang2021explanations, yang2020visual} which means the human ignores the AI's prediction though the AI is correct; (3) appropriate reliance \cite{poursabzi2021manipulating, wang2021explanations, yang2020visual} which means an ideal situation where the human can adopt the AI's suggestion when the AI makes a correct prediction and vice versa. For example, Wang et al. \cite{wang2021explanations} evaluate the effects on trust calibration by measuring users' awareness of model uncertainty and calculating the appropriate trust ratio, overtrust ratio, and undertrust ratio in the given instances.

\subsubsection{Fairness Judgement}

Studying how people perceive the fairness of AI and what design impacts the perception is an active research area. These studies primarily rely on subjective metrics, from general perceived fairness \cite{anik2021data, dodge2019explaining, green2019disparate, harrison2020empirical, van2021effect} to perceptions of more fine-grained types of fairness such as individual fairness \cite{lee2019procedural}, group fairness \cite{lee2019procedural}, process fairness \cite{binns2018s}, deserved outcome \cite{binns2018s}, feature fairness \cite{binns2018s, van2021effect}, and accountability (i.e., the extent to which participants think the system is fair and they can control the outputs the system produces) \cite{rader2018explanations}.
For example, Anik et al. \cite{anik2021data} use a subjective questionnaire to measure users' perceived fairness of the system and the AI training. In a recidivism prediction task, Dodge et al. \cite{dodge2019explaining} design a questionnaire to subjectively measure users' perception of fairness.
Only a small number of studies leveraged decision bias (e.g., the action to follow the model’s recommendations despite their lack of fairness) \cite{green2019disparate, green2019principles} as an objective metric of perceived fairness.

\subsubsection{Other factors.}
One important factor that end-users are always concerned about is the privacy of their data \cite{mohseni2021multidisciplinary, liao2021human, arrieta2020explainable}. An explanation is expected to facilitate privacy checking for users. Another factor is the bias in the model. A model can be biased either due to the biased training data or the biased feature engineering \cite{arrieta2020explainable}. Explanation, especially the explanation of the training data can help deal with this problem. However, there is little explicit empirical research on these topics.

\renewcommand{\arraystretch}{1.5}
\begin{table}[tp]  
  \centering  
  \fontsize{8}{8}\selectfont

\begin{threeparttable}  

\caption{Evaluation related to users' perceptions of AI}\label{table:posttask-questionnaire-student} 
\begin{tabular}{p{2cm}p{1.5cm}p{10cm}}
%\begin{tabular}{cccccc}

\toprule
\textbf{Evaluation Goal}&\textbf{Subjectivity}&\textbf{Metrics}\\ \hline
\textbf{G1: Understanding/Mental Model}&subjective&Self-reported understanding \cite{anik2021data, buccinca2020proxy, cai2019effects, cheng2019explaining, wang2021explanations, yang2020visual}, confidence in simulation \cite{alqaraawi2020evaluating, nguyen2018comparing}, intuitiveness \cite{szymanski2021visual}, confidence in understanding \cite{kulesza2012tell}, perceived transparency/interpretability \cite{rader2018explanations, tsai2021exploring}, ease of understanding \cite{guo2019visualizing, poursabzi2021manipulating}\\\cline{2-3}

&objective&Forward simulation \cite{abdul2020cogam, buccinca2020proxy, chromik2021think, nourani2021anchoring, poursabzi2021manipulating, ribeiro2018anchors, wang2021explanations, yu2019trust}, counterfactual simulation \cite{hase2020evaluating, wang2021explanations}, model errors detection \cite{wang2021explanations, poursabzi2021manipulating}, identifying important features \cite{cheng2019explaining, szymanski2021visual}, correctness of estimated model performance/accuracy \cite{nourani2021anchoring, smith2020no}, comprehension quiz \cite{cheng2019explaining, gero2020mental, kulesza2012tell, wang2021explanations}, correctness of described model behaviors \cite{chromik2021think, kulesza2013too, rader2018explanations}\\\hline

\textbf{G2: Trust and Reliance}&subjective&Self-reported trust \cite{abdul2020cogam, buccinca2020proxy, cheng2019explaining, chromik2021think, poursabzi2021manipulating, springer2019progressive, tsai2021exploring}, model confidence/acceptance \cite{alqaraawi2020evaluating, chromik2021think, szymanski2021visual, wang2021explanations}, self-reported agreement/reliance \cite{chandrasekaran2018explanations}, perceived accuracy \cite{kocielnik2019will, smith2020no, springer2019progressive}\\\cline{2-3}

&objective&Agreement/acceptance of model suggestions \cite{bansal2021does, lai2020chicago, lai2019human, liu2021understanding, wang2021explanations, yin2019understanding, zhang2020effect}, switch \cite{yin2019understanding, zhang2020effect}, weight of advice \cite{poursabzi2021manipulating}, disagreement/deviation \cite{poursabzi2021manipulating}, choice to use the model \cite{ribeiro2016should}\\\hline

\textbf{G3: Trust Calibration}&objective&over-reliance \cite{buccinca2021trust, bussone2015role, wang2021explanations, yang2020visual}, under-reliance \cite{bussone2015role, wang2021explanations, yang2020visual}, appropriate reliance \cite{poursabzi2021manipulating, wang2021explanations, yang2020visual}\\\hline

\textbf{G4: Fairness Judgement}&subjective&Perceived fairness \cite{anik2021data, dodge2019explaining, green2019disparate, harrison2020empirical, van2021effect}, individual/group fairness \cite{lee2019procedural}, deserved outcome \cite{binns2018s}, feature fairness \cite{binns2018s, van2021effect}, accountability \cite{rader2018explanations}\\\cline{2-3}
&objective&Decision bias \cite{green2019disparate, green2019principles}\\

% \textbf{G9: Other human-centered goals}&&\\\hline

\bottomrule
\end{tabular}
\end{threeparttable}  

\end{table}

\subsection{Evaluation Related to Users' Interaction Experience.}
The explanation design not only affects users' perceptions of the AI itself but also affects users' experience when interacting with the explanatory system. In the user experience (UX) domain, there are rich evaluation perspectives. However, in XAI empirical studies, we mainly divide user interaction experience from three perspectives, satisfaction/perception of the system, cognitive load/mental demand, and engagement, as shown in Table 4.

\subsubsection{Satisfaction/Perception of the system.}

Subjective metrics are often used, such as users' satisfaction with the AI \cite{kocielnik2019will, lage2019evaluation, lucic2020does, tsai2021exploring}, perceived helpfulness/usefulness \cite{buccinca2020proxy, cai2019human, kocielnik2019will, yang2020visual}, effectiveness \cite{tsai2021exploring}, quality \cite{tsai2021exploring}, appropriateness \cite{buccinca2020proxy}, preference \cite{kulesza2012tell, lee2020co, ribeiro2018anchors, yang2020visual}, etc. Besides, some studies measure system complexity \cite{buccinca2021trust}, ease of use \cite{anik2021data, tsai2021exploring}, system frustration \cite{kocielnik2019will, lee2020co, smith2020no}, information richness \cite{anik2021data, lee2020co, lee2021human}, learning effect \cite{tsai2021exploring}, etc. Focusing on explanation, people’s perceived explanation quality \cite{kulesza2013too, kunkel2019let, szymanski2021visual}, explanation usefulness \cite{cai2019human, lee2021human, nourani2021anchoring, szymanski2021visual}, easiness to use explanation \cite{szymanski2021visual} are often subjectively measured.

For example, in an average reading time of a magazine article prediction task, Szymanski et al. \cite{szymanski2021visual} measure users' perception of ease-of-use of each type of explanation. In a UX problem finding task, Fan et al. \cite{fan2021human} use a subjective questionnaire to measure users' satisfaction. And in a vocal emotion recognition task, Zhang et al. \cite{zhang2021towards} measure users' perceived system helpfulness on a 7-point Likert scale. In a sentiment classification and a QA task, Bansal et al. \cite{bansal2021does} use a subjective questionnaire to measure users' perceived helpfulness. In a recidivism prediction task, Liu et al. \cite{liu2021understanding} use a subjective questionnaire to measure users' perception of AI assistance’s usefulness. 
% In a virtual agent environment, Ehsan et al. \cite{ehsan2021explainable} use a subjective questionnaire to measure users' perceived intelligence and friendliness of different agents with different explanations.

% Of the papers we surveyed, only Cai et al. \cite{cai2019effects} leveraged objective satisfaction measures, using the time spent with the AI system to reflect users’ interest and satisfaction.

% Researchers have asked participants about their subjective satisfaction with
% the process \cite{dietvorst2018overcoming}, confidence in the process \cite{dietvorst2018overcoming}, frustration/annoyance \cite{kulesza2013too, smith2020no}, mental demand/effort \cite{buccinca2020proxy, buccinca2021trust, kulesza2013too, weerts2019human}, workload \cite{cai2019human, lee2021human}.

% UX

\subsubsection{Cognitive Load/Mental Demand}
On the one hand, the explanation can help users understand the AI. On the other hand, since the explanation contains extra information about the model, it might lead to cognitive load or much mental demand for users. To measure the effects of explanations, researchers have asked participants about their mental demand/effort \cite{buccinca2020proxy, buccinca2021trust, kulesza2013too, weerts2019human}, workload \cite{cai2019human, lee2021human}. For example, in a task where users need to replace the highest calorie ingredient in a dish with another lower calorie but similar taste ingredients, Buçinca et al. \cite{buccinca2021trust} use a subjective questionnaire to measure users' mental demand and users' perception of system complexity. In a COVID-19 self-diagnosis task, Tsai et al. \cite{tsai2021exploring} use the NASA-TLX scale to measure users' cognitive load. In a house price prediction task, Abdul et al. \cite{abdul2020cogam} use a mixed measure to evaluate users' cognitive load, including the reading time, users' self-reported cognitive load, recall reconstruction score, and recognition score. In a video game event recall task, Robertson et al. \cite{robertson2021wait} use a subjective questionnaire to measure users' cognitive effort.

\subsubsection{Engagement}
Engagement plays a crucial role in human-computer interaction. Some works also measure the effects of explanation on users' engagement. For example, in a UX problem finding task, Fan et al. \cite{fan2021human} use the interaction logs on the interface to measure users' engagement.

\renewcommand{\arraystretch}{1.5}
\begin{table}[tp]  
  \centering  
  \fontsize{8}{8}\selectfont

\begin{threeparttable}  

\caption{Evaluation related to users' interaction experience.}\label{table:posttask-questionnaire-student} 
\begin{tabular}{p{2cm}p{1.5cm}p{10cm}}
%\begin{tabular}{cccccc}

\toprule
\textbf{Evaluation Goal}&\textbf{Subjectivity}&\textbf{Metrics}\\ \hline

\textbf{G5: System Satisfaction/Usability}&subjective&Satisfaction \cite{biran2017human, dietvorst2018overcoming, kocielnik2019will, lage2019evaluation, tsai2021exploring}, helpfulness/usefulness \cite{buccinca2020proxy, cai2019human, kocielnik2019will, yang2020visual}, effectiveness \cite{tsai2021exploring}, quality \cite{tsai2021exploring}, appropriateness \cite{buccinca2020proxy}, preference \cite{kulesza2012tell, lee2020co, yang2020visual}, complexity \cite{buccinca2021trust}, ease of use \cite{anik2021data, tsai2021exploring}, system frustration \cite{lee2020co, smith2020no}, richness/informativeness \cite{anik2021data, lee2021human}, learning \cite{tsai2021exploring}, recommendation to others \cite{kocielnik2019will}, usefulness/helpfulness of explanation \cite{cai2019human, nourani2021anchoring, szymanski2021visual}, easiness to use explanation \cite{szymanski2021visual}, quality/soundness/completeness of explanation \cite{szymanski2021visual}\\\hline

\textbf{G6: Cognitive Load/Mental Demand}&subjective&mental demand/effort \cite{buccinca2020proxy, buccinca2021trust, kulesza2013too, weerts2019human}, workload \cite{cai2019human, lee2021human}\\\hline

\textbf{G7: Engagement}&objective&Browsing time on the interface \cite{fan2021human}\\

\bottomrule
\end{tabular}
\end{threeparttable}  

\end{table}

\subsection{Evaluation Related to Task}
For the task-related measures, existing works mainly focus on two aspects, task performance, and task efficiency.

\subsubsection{Task Performance}
Explanations are often designed to assist users to perform tasks, thus task performance is a popular measurement in existing empirical studies.

Task performance is usually evaluated by objective measures. In classification tasks, the most commonly used metric is accuracy \cite{bansal2019beyond, dressel2018accuracy, ghai2021explainable, lage2019evaluation, lai2020chicago, lai2019human, nguyen2018comparing}. Researchers are interested in comparing the accuracy of the human-AI team with the accuracy of human-alone or AI-alone. 
For example, in a food percent fat content of nutrients recognition task, Buçinca et al. \cite{buccinca2020proxy} calculate the percentage of correct answers to measure the task performance. In an income prediction task, Zhang et al. \cite{zhang2020effect} use accuracy to measure performance. In a cooking video activity recognition task, Nourani et al. \cite{nourani2021anchoring} use the number of errors to measure the task performance.

Besides accuracy, other metrics in classification tasks are also used to evaluate the task performance, such as F1 \cite{biran2017human}, precision \cite{biran2017human}, recall \cite{biran2017human}, AUC-ROC \cite{dressel2018accuracy}, false positives rate \cite{carton2020feature, dressel2018accuracy, green2019disparate}, false negatives rate \cite{carton2020feature, dressel2018accuracy}, true positive rate \cite{dressel2018accuracy}, true negative rate \cite{dressel2018accuracy}, etc. For example, in a UX problem-finding task, Fan et al. \cite{fan2021human} use accuracy, precision, and recall of UX problem detection to measure task performance.

In addition to objective metrics, subjective metrics are used to measure human perception of task performance, such as the perceived accuracy (i.e., self-rated error/accuracy) \cite{dietvorst2018overcoming, lai2019human, tsai2021exploring}, humans' confidence in the decisions \cite{green2019disparate, green2019principles, guo2019visualizing}. Compared to objective measures, subjective metrics are less intuitive and used.

%task performance

\subsubsection{Time Spent/Task Efficiency}

Another task-related dimension is efficiency, which means how efficiently the user can complete the task with the AI's assistance. The most common objective metric is time spent on the task \cite{abdul2020cogam, carton2020feature, cheng2019explaining, slack2019assessing, gero2020mental, kocielnik2019will, lage2019evaluation, smith2020no, yang2020visual}. For example, in a vocal emotion recognition task, Zhang et al. \cite{zhang2021towards} measure the efficiency by the logged task times for different pages. In a graduate admission prediction task, Cheng et al. \cite{cheng2019explaining} collect users' spent time when interacting with different explanatory interfaces.

\renewcommand{\arraystretch}{1.5}
\begin{table}[tp]  
  \centering  
  \fontsize{8}{8}\selectfont

\begin{threeparttable}  

\caption{Evaluation related to the task}\label{table:posttask-questionnaire-student} 
\begin{tabular}{p{2cm}p{1.5cm}p{10cm}}
%\begin{tabular}{cccccc}

\toprule
\textbf{Evaluation Goal}&\textbf{Subjectivity}&\textbf{Metrics}\\ \hline

\textbf{G8: Task Performance}&subjective&Self-rated error/accuracy \cite{dietvorst2018overcoming, lai2019human, tsai2021exploring}, confidence in the decisions \cite{green2019disparate, green2019principles, guo2019visualizing}\\\cline{2-3}
&objective&Accuracy/error \cite{bansal2019beyond, ghai2021explainable, lage2019evaluation, lai2020chicago, lai2019human}, precision \cite{fan2021human}, recall \cite{fan2021human, narayanan2018humans}, false positive rate \cite{carton2020feature, dressel2018accuracy, green2019disparate}, false negative rate \cite{carton2020feature, dressel2018accuracy}, mean prediction error \cite{poursabzi2021manipulating}\\\hline

\textbf{G9: Time Spent/Efficiency}&objective&time taken on the task \cite{abdul2020cogam, arshad2015investigating, carton2020feature, cheng2019explaining, slack2019assessing, gero2020mental, kocielnik2019will, lage2019evaluation, smith2020no, yang2020visual}\\

\bottomrule
\end{tabular}
\end{threeparttable}  

\end{table}

\subsection{Summary and Thinking}
In this section, we review the commonly used evaluation metrics in empirical studies.

\subsubsection{Challenges}
\textbf{Lack of standard evaluation system}
From the surveyed paper, we find that existing empirical studies have not yet formed a unified set of evaluation methods. Each work has adopted different evaluation methods from each other, which hinders the collaborative progress of the field and the mutual utilization of results. Many studies even directly use evaluation methods designed by themselves. Such a phenomenon is mainly due to the fact that the empirical research on XAI is still in the preliminary stage, and a standard evaluation system has not yet been formed. 

\subsubsection{Future research opportunities}
\textbf{Developing an evaluation system for XAI empirical studies.}
We believe that the next step in this field is to form a standard evaluation system. First, it requires a detailed division of different XAI tasks, user categories, experiment scales, goals, etc. In Section 5, we only divide the evaluation methods based on evaluation goals, which is not enough for an evaluation system because a number of other factors need to be considered. Second, the evaluation system must be scalable, because the research on XAI is just at a young age, and the existing evaluation methods are only some preliminary exploration. Therefore, there will be many excellent and appropriate evaluation methods for each kind of XAI in the future, thus the system needs to flexibly absorb new methods, and discard or improve the old methods. Finally, to build such a system, all researchers need to work together, and some open-source and collaborative methods can be considered to maintain the evaluation system.

%% file: sections/findings.tex
\section{Findings and Pitfalls Derived from Empirical Studies}

From carefully organized user studies, the empirical research of XAI has obtained valuable findings from the results. These findings can help researchers or practitioners understand the possible effect of different XAI designs on a specific user group in a specific task scenario, and help researchers get references when designing new explanations or new empirical studies. However, we find that in the previous surveys on XAI, there is scarce work to summarize the relevant literature from the perspective of findings. There are two possible reasons. First, existing XAI surveys do not focus on the perspective of empirical study \cite{hohman2018visual, mohseni2021multidisciplinary, tomsett2018interpretable, bhatt2020explainable, hong2020human}. Second, current works on XAI empirical study often focus on a specific task, propose a specific (set of) explanation design, verify it on a specific type of user, and use a specific evaluation method. These make it hard to make a fair comparison between the findings obtained from the different studies. We acknowledge that the findings obtained in different experimental settings may not be generalized to other experimental settings. However, we believe that analysis and summarization of the findings will help readers have a deeper understanding of the XAI design mentioned earlier in this survey.

In order to understand the current findings in the XAI empirical studies from an overall perspective, we use the qualitative analysis method commonly used in HCI to conduct a thematic analysis of the findings in the literature and extract some commonalities. We believe that the themes in these findings are of great value to researchers and practitioners, such as establishing an overall understanding of the existing work, building more comprehensive and reasonable expectations for their own experimental design, and being aware of possible pitfalls and failures, controlling confounding design factors. Overall, we categorize these findings into common findings and pitfalls, and the latter focuses on the current negative effects of XAI design. Next, we will summarize the findings based on themes. Under each theme, we will list several representative empirical studies as examples. Note that we do not list ``well-known positive'' findings that support the effectiveness of the explanation, such as appropriate XAI design can help users understand the model or can improve users' satisfaction with the AI system.

\subsection{Common findings}

\subsubsection{A moderate granularity of explanation could benefit users’ understanding of AI}
There exists a trade-off between complex detailed explanations and simple explanations. For example, Mishra et al. \cite{mishra2021crowdsourcing} find that a balance between coarse and fine-grained explanations lead to better users mental model of the model's predictions. Abdul et al. \cite{abdul2020cogam} correlate the complexity of visual explanations with the cognitive load of the user, and propose a method COGAM to trade-off between cognitive load and accuracy.

Apart from designing a moderate granularity of explanation, some empirical studies suggest a progressive disclosure of explanations. For example, Springer et al. \cite{springer2019progressive} find that users can benefit from progressive disclosure of explanations where simplified feedback that helps users build heuristics about the system is shown initially. Similarly, Nourani et al.’s \cite{nourani2021anchoring} study suggests that before using the detailed instance-level explanations, a higher-level explanation could help users build more accurate mental models.

\subsubsection{Interactive explanation may and may not increase users’ understanding or task performance.}

The interactive explanation has been proposed in visualization and HCI domains, which is expected to enhance users’ exploration and thus understanding of the AI model. However, through empirical studies, there are mixed findings of the effects of interactive explanation on users’ understanding.

On the one hand, Cheng et al. \cite{cheng2019explaining} compare the effects of interactive explanations on users’ understanding, trust, and time spent. They conduct the user study on a graduate student admission prediction task via 199 AMTurk participants and find that compared to the static interfaces, interactive interfaces increase the understanding of the algorithm. For data scientists, Narkar et al. \cite{narkar2021model} propose an interactive explanatory tool for multi-level model comparison in autoML, and find that participants gave high ratings to the usability of this tool. On the contrary, in a crowdsourcing user study with two types of tasks, recidivism prediction, and profession prediction. Liu et al. \cite{liu2021understanding} try to understand the effect of interactivity of explanation and data distribution on human-AI decision making. They find that interactive explanations may reinforce human biases and lead to limited performance improvement.

\subsubsection{Expertise matters how users perceive an explanation.}

It has been widely recognized that users’ expertise will play a critical role in how users perceive an explanation and the effectiveness of an explanation design \cite{suresh2021beyond, mohseni2021multidisciplinary, ehsan2021explainable, schaffer2019can, anik2021data, szymanski2021visual, yang2020visual}. The expertise includes both AI expertise and domain expertise \cite{suresh2021beyond, mohseni2021multidisciplinary}. Usually, experts users could gain more from explanations than novices. For example, Szymanski et al. \cite{szymanski2021visual} find that AI novices gain less benefit from explanations but are also more likely to have illusory satisfaction compared to experts. Fan et al. \cite{fan2021human} suggest that compared with non-experts, such as crowdworkers, expert users tend to have higher confidence in their judgments and might adopt different strategies in human-AI collaboration \cite{fan2019vista, soure2021coux}. For a decision-making scenario, Wang et al. \cite{wang2021explanations} compare different types of explanations in AI-assisted decision-making. They conduct a user study on two types of tasks, forest cover prediction (users lack the expertise) and recidivism prediction (users have the expertise). They find that users' expertise in different decision-making tasks highly influences the effectiveness of different XAI methods.

\subsubsection{Apart from expertise, users’ other intrinsic characteristics can affect their perceptions of explanations.}

Recent empirical studies find that there are some other users’ characteristics that could affect the effects of an explanation, such as users' general trust in AI \cite{dodge2019explaining}, locus of control \cite{rotter1966generalized}, visual literacy \cite{abdul2020cogam, boy2014principled}), cognitive load disposition \cite{ghai2021explainable}, need for cognition \cite{cacioppo1984efficient} (a personality trait reflecting one’s general motivation to engage in effortful mental activities, such as thinking). Also, Anik et al. \cite{anik2021data} suggest that users' perceptions can be influenced by their prior exposure to the AI concepts, such as what they have seen in the media. Also, humans’ prior intuition is probably an informative distinguishing factor that can affect the human perception and interpretation of the explanations \cite{lai2019human}.

Taking the need for cognition as an example, some studies suggest that people are less able to process explanations effectively if the time and cognitive resources are constrained \cite{robertson2021wait, xie2020chexplain}. Ghai et al. \cite{ghai2021explainable} show people who have a low score in need for cognition will be less satisfied when adding explanations in an active learning setting. To reduce users’ overreliance on AI, Buçinca et al. \cite{buccinca2021trust} propose three kinds of cognitive forcing functions in a crowdsourcing study. Their results show that the proposed cognitive forcing interventions benefit participants who get a higher score in need for cognition more. Besides the cognitive factor, Dodge et al. \cite{dodge2019explaining} conduct an empirical study to explore how explanations affect users’ fairness judgment. Through a crowdsourcing user study of a recidivism prediction task, they find that how users react to different styles of explanation will be influenced by some individual differences, such as their prior positions and judgment criteria of algorithmic fairness.

\subsubsection{Different types of explanation could benefit different design goals.}

A large number of empirical studies have been designed and conducted to understand how users perceive, process, and use different types of explanation \cite{wang2021explanations, zhang2021towards, lai2020chicago, tsai2021exploring, narkar2021model, yang2020visual, dodge2019explaining}. Generally, in different scenarios, for different objectives, one type of explanation could be more effective than another. For example, Dodge et al. \cite{dodge2019explaining} explore the effect of different types of explanations on calibrating users' perceived fairness of ML models. They find that compared to global explanations, local explanations are more effective in calibrating users’ fairness judgment of ML models, because local explanations can highlight unfair features used for individual predictions. In a different setting, Tsai et al. \cite{tsai2021exploring} design three types of explanations for online symptom checkers. Through a lab-controlled user study, they find that users may have different preferences for different explanations. Static or ``one-size-fits-all'' explanations cannot fulfill users' needs as they would like to get some control and customization in what explanations are conveyed to them. These findings imply that the explanations are expected to be adaptive and controllable based on the users’ real-time information needs. 

% Comparing various types of explanations against one another to understand which method works better against certain biases, or incorporating multiple explanation scopes within one interface might allow users to decide what they want to explore to understand the model decisions better.

\subsubsection{Tutorial and introduction of the model can benefit users’ understanding.}

Sometimes, to achieve an effective interaction with AI, users need to access the global information of an AI system to build an appropriate understanding of the underlying model and data. Recently, some works provide documentation to present global information to users (e.g., Model cards \cite{mitchell2019model}, Datasheet \cite{gebru2021datasheets}, FactSheets \cite{arnold2019factsheets}). Empirical studies have shown that offering users a tutorial or an introduction to the whole model or data can lead to a more accurate mental model. For example, Lai et al. \cite{lai2020chicago} propose some types of tutorials of ML models, and compare them with no-tutorial conditions in a deceptive review detection task with MTurk participants. They find that tutorials can indeed improve human performance to some extent.

\subsection{Pitfalls}

\subsubsection{Transparent systems do not necessarily lead to high trust or better task performance.}

Some explanations are designed to improve users’ trust and task performance based on an assumption that a transparent system will help users understand the AI and thus improve users’ trust and task outcomes. However, several studies find that explanations could lead to some negative effects, such as decreasing users' situation awareness and leading to worse task performance \cite{zhang2020effect, cummings2004automation, poursabzi2021manipulating, schaffer2019can}, because explanations could overwhelm or overload users with much information about the system.

For example, Robertson et al. \cite{robertson2021wait} investigate different behavior explanations for a real-time strategy game setting and find that a \emph{why} explanation does not improve users’ task performance because users lack enough cognitive resources to interpret the \emph{why} answers in real-time. In a student admission prediction task, Cheng et al. \cite{cheng2019explaining} find that showing the explanation does not increase users' trust in the algorithm compared with a black-box model without explanation. Bansal et al. \cite{bansal2021does} design an empirical study to investigate the effects of explanations on human-AI team performance. They find that explanations do not increase team performance but increase the chance that users follow the AI’s suggestion, regardless of its correctness. And they suggest that ``instead of convincing, explanations should be informative''. Besides, some recent empirical studies investigate the effects of the level of transparency on users' trust in the AI system \cite{cheng2019explaining, kunkel2019let, poursabzi2021manipulating, schaffer2019can}. They manipulate the model configuration to increase the transparency of the model by providing explanations, reducing the feature number, and allowing users to inspect the model behavior, using a white-box model. However, they find no significant improvement in users' trust. For example, in an apartment price prediction task, they do not find that participants follow the transparent model's predictions when the model makes a correct prediction. Furthermore, showing participants a clear model hinders their ability to detect and correct the mistakes of the model, which is possibly due to information overload.

\subsubsection{Human-AI teams with comparable performance do not guarantee complementary performance.}

A common XAI scenarios is decision-making, where the human makes the final decisions based on the AI's suggestions and explanations. Hence, the human-AI team's performance, i.e., the accuracy of the final decision, is regarded as an essential factor to measure the effectiveness of the XAI design. In recent works, researchers begin to investigate whether the human-AI team can achieve a complementary performance when the human and AI have comparable performance. However, some evidence shows that comparable performance alone cannot guarantee complementary performance \cite{zhang2020effect, rastogi2020deciding, liu2021understanding, bansal2021does}.

For example, from a crowdsourcing study on an income prediction task, Zhang et al. \cite{zhang2020effect} find that although a confidence score can help calibrate people’s trust in an AI model, it is still not sufficient to improve team performance. The joint team performance may also depend on whether the human and AI have non-overlapping knowledge and different error boundary. In a similar view, Bansal et al. \cite{bansal2021does} emphasize that what plays an important role is the complementary knowledge between humans and AI. Rather, in an ideal team, humans and AI could maximize their talents in different dimensions. For example, for clinical decision-making, AI could quickly sort similar cases from the database based on its computing power, and the doctor could make a unique diagnosis based on her rich domain knowledge \cite{tonekaboni2019clinicians}.

\subsubsection{Too technical or visual-complicated explanations might be hard to understand, even for experts.}
Although explanations are usually designed for users' understanding. However, some explanation methods are technique-centered rather than experience-centered \cite{liao2021human}, which hinders users' ability to correctly or fully understand them. The information contained in sophisticated explanations can also lead to information overload and prevent people from forming a proper mental model \cite{springer2019progressive}. For example, Zhang et al. \cite{zhang2021towards} propose a framework to generate different types of relatable explanations and in their study, they find that saliency visualization is not useful since it is too technical and complicated. Yang et al. \cite{yang2020visual} focus on visual explanations and explore the effects of spatial layout and visual representation. They find that visual explanations can lead to inappropriate trust if an explanation is difficult to understand. Generally, expert users have a stronger ability to analyze complex explanations. For example, Szymanski et al. \cite{szymanski2021visual} find that although lay users prefer visual explanations, they get significantly worse performance with it. And they find experts can better understand visual and textual explanations. However, it should be noted that prior research shows that even ML experts face challenges to interpret them correctly without assistance \cite{kaur2020interpreting, chromik2021think}.

% Suggested by literature, to seek technical and design solutions that reduce the cognitive workload imposed by XAI, designers can reduce the quantity and improve the consumability of information. For example, studies suggest that muti-modalities (text, visual, audio, etc.) can be leveraged to aid attention and understanding of XAI \cite{robertson2021wait, szymanski2021visual}. Progressive disclosure \cite{springer2019progressive}, starting with simplified or high-level transparency information and revealing details later or upon user requests, is another effective approach to reduce cognitive workload.

% For a movie recommendation chatbot, Wilkinson et al. \cite{wilkinson2021or} design “why” and “why not” justifications. They find that not everyone prefers the visual explanations, and there are both experts and lay users preferring textual explanations because they find the visual explanations to be overwhelming.

\subsubsection{Cognitive biases can affect users’ mental model of XAI systems and can lead to negative effects.}

Based on cognitive science, some empirical studies have looked into how cognitive biases exist and affect users’ perceptions of an explanation. And they reveal that cognitive biases can prevent XAI from doing what it was designed to do. For example, Nourani et al. \cite{nourani2021anchoring} investigate the effects of anchoring bias on users’ mental model formation. And they find that the first impression plays a key role in users' perception of AI. Specifically, users who encounter system strengths early are more prone to automation bias and make significantly more errors due to positive first impressions. On the contrary, users who observe system weaknesses early make significantly fewer errors because they tend to rely more on themselves, while the negative first impression also makes them underestimate the capability of the model. Chromik et al. \cite{chromik2021think} compare a moderated interaction with AI and an unmoderated interaction with AI, and explore the effects of lay users’ perceived understanding of the AI. They find that in the unmoderated setting, participants often adopt heuristic thinking and cannot realize the incompleteness of their understandings until they see their test results.

% There is also an empirical study showing that proxy tasks can be misleading. Buçinca et al. \cite{buccinca2020proxy} find that using proxy tasks (e.g., simulation test) to evaluate explainable AI systems may lead to different results than testing the systems on actual decision-making tasks. To explain their results, the authors hypothesized that proxy tasks artificially forced participants to pay attention to the AI, but when participants were presented with actual decision-making tasks, they focused on making the decisions and allocated fewer cognitive resources to analyzing AI-generated suggestions and explanations. This reason can also be applied to think-aloud. They find the use of the think-aloud method — a standard technique for evaluating interactive systems — can also substantially impact how participants allocate their mental effort. Because participants were asked to think aloud, they suspect that they exerted additional cognitive effort to engage with the explanations and analyze the reasoning behind their decisions.

Based on theories of cognitive science and psychology, researchers have examined humans' cognitive processes when they interpret AI explanations. One of the most representative theories mentioned in existing empirical studies is the dual-process of cognition \cite{buccinca2020proxy}. From the dual-process theory \cite{kahneman2011thinking, cacioppo1984elaboration, wason1974dual}, humans' cognitive processes can be driven by two systems: System 1 and System 2. The former makes humans think fast and process information in an automatic manner, whereas the latter makes humans think slow and engaged in deliberative and analytical thinking. System 1 usually relies on developed heuristics (rules-of-thumb or mental short-cuts) which could lead to cognitive biases if applied inappropriately \cite{kahneman2011thinking}. Based on this theoretical foundation, there is an increasing awareness \cite{buccinca2020proxy, ehsan2021explainable, nourani2021anchoring, rastogi2020deciding, wang2019designing} that while XAI developers assume that humans will carefully digest every information in explanations (analytic System 2 thinking), in reality, people are more likely to adopt System 1 thinking as it is fast and labor-saving. Cognitive biases can lead to many negative consequences such as over-trust in XAI (mentioned below), which can be attributed to users' associating explanations with AI competence in System 1 heuristic \cite{chromik2021think, nourani2021anchoring}. 

% One way to mitigate these biases would be to use Cognitive Forcing Functions (CFFs)–interventions that disrupt heuristic reasoning and promote System 2 analytical thinking \cite{buccinca2021trust, rastogi2020deciding}.

% For example, to mitigate users' cognitive biases, Buçinca et al. \cite{buccinca2021trust} design three kinds of cognitive forcing functions, 1) asking the person to make a decision before seeing the AI’s recommendation, 2) slowing down the process, and 3) letting the person choose whether and when to see the AI recommendation. The results demonstrate that cognitive forcing significantly reduced overreliance compared to the simple explainable AI approaches.
% In another study, to mitigate cognitive biases, Rastogi et al. \cite{rastogi2020deciding} propose an adaptive time allocation strategy based on model’s confidence. Specifically, when the model is in high confidence, the system gives users less time to think, and vice versa. They find that the time allocation strategy with explanation can effectively de-anchor the human and improve collaborative performance when the AI model has low confidence and is incorrect. This also shows that cognitive biases can have a negative effect on the effectiveness of explanation design.

\subsubsection{Explanation can lead to users’ over-trust even the model or the explanation is not “good”.}

Some recent works show that explanations, even if they are placebic or randomly generated, may improve humans' trust in AI predictions \cite{bansal2021does, green2019disparate, green2019principles, lai2019human}. For example, Kaur et al. \cite{kaur2020interpreting} find that the existence of explanations can make data scientists hold an over-confidence and mistakenly think that the model is ready for deployment. In addition, there is the concern of illusory understanding which means that humans can over-estimate their understanding gained from XAI \cite{chromik2021think}. For the underlying mechanism of why XAI could lead to over-reliance, some studies find that explanations are generally interpreted as a signal of competence, no matter what content is in the explanation, and just the presence of an explanation can increase humans' trust in the AI \cite{bansal2021does, ghai2021explainable, poursabzi2021manipulating, zhang2020effect}. As mentioned in the above section, humans are easily engaged in System 1 thinking so that they simply associate explanation with AI's capability without engaging analytically with the model behaviors.

Besides the explanation, other types of explanatory elements could also lead to over-trust issues. For example, Lai et al. \cite{lai2019human} find that participants are more likely to trust models with accuracy statements than models without accuracy statements, even if poor performance is stated. These observations are consistent with prior work on numeracy which suggests that it is hard for humans to interpret and act on numbers \cite{berwick1981doctors, peters2006numeracy, reyna2008numeracy, slovic2006risk}. Also, Lai et al. \cite{lai2019human} show that adding random heatmap as explanations can enhance humans' trust. Such an over-reliance phenomenon does not just happen to non-expert users. Some papers find that both expert users and non-expert users have unwarranted faith in numbers. For example, Ehsan et al. \cite{ehsan2021explainable} find that participants in both AI-expert and AI-novice groups have unwarranted faith in numbers. To mitigate these issues, recent studies find that presenting information about uncertainty is an effective means to help users maintain appropriate trust in AI \cite{bansal2021does, zhang2020effect, rastogi2020deciding}.

\subsection{Summary and Thinking}
In this section, we summarize the common findings and pitfalls of current XAI design, which can offer valuable implications for researchers to develop new XAI interfaces and conduct new empirical studies.
\subsubsection{Challenges}
\textbf{The experimental conclusions are diverse and difficult to generalize}
Although we have summarized some common experimental findings in Section 6, we find that diversity exists in existing experimental findings. For example, some studies find that interactive XAI design helps to enhance users' trust in AI, while some works find that interactive XAI does not have such an effect. There are many reasons behind this, such as different XAI design details, different users, different tasks, different experimental procedures, different AI algorithms, different evaluation metrics, and more. Although the diversity of research results is beneficial for the healthy development of the young field, it prevents the generalization of different studies and hinders the joint efforts of different research. However, we suggest that when different works analyze their experimental conclusions, they can generalize the experimental conclusions to a more general level by analyzing the reasons and patterns behind them based on relevant theories, which will help researchers to establish a systematic understanding in this field.

\textbf{Lack of successful application cases}
Although the explanation of recommender systems has been widely adopted, and many products now have a simple explanation of the newly added AI functions, we still rarely see the success of other AI tools in our lives, such as a decision support system. On the one hand, this stems from the lack of attention to users' needs for interpretability. Many AI products still only focus on providing functions and do not pay attention to users' needs for interpretation of the functions provided. On the other hand, providing any explanation will actually bring extra information to the user, which will affect the existing task flow, and may also be inconsistent with the mental model that the user has established before. If not well designed, the steps in which the user interacts with the explanation interface may degrade the user's user experience, and even the explanation provided does not really help the user understand the system. In short, in order to promote the application of XAI in real life and work, the cooperation between academia and industry, as well as the cooperation of multidisciplinary researchers is needed.

\subsubsection{Future research opportunities}
\textbf{Developing more theory-grounded XAI interface and empirical studies.} From the analysis of existing studies, we find many findings can echo well-known theories from social science, psychology, cognitive science. Thus, we may leverage explanation-related theories, such as how a human explains to another human, to guide more effective XAI design and more scientific empirical study.

\textbf{Designing more effective XAI methods for human-AI decision-making.} As mentioned in this section, XAI has the potential to help users build more accurate mental models of the AI and achieve complementary human-AI team performance in decision-making. However, we also notice that the current explanation design is far from successful due to a lack of consideration of humans' cognitive processes. In the future, we can focus on specific decision-making tasks, and leverage XAI to mitigate the potential cognitive biases.

% Develop frameworks to characterize decision tasks
% Justify choices of decision task
% Expand datasets availability

% Human-centered analysis to define the design space of AI assistance elements
% Extend the design space and studies beyond decision trials
% Task-driven studies to complement feature-driven studies

% Make choices of evaluation metrics by research questions/hypotheses and targeted constructs
% Work towards common metrics and a shared understanding on the meanings, strengths and weaknesses of different evaluation methods
% Keep reflecting on common evaluation metrics as value-laden choices

%% file: sections/framework.tex
\section{A framework for Human-Centered Design of XAI with Empirical Studies}

\begin{figure*}[h]
	\centering 
	\includegraphics[width=0.9\linewidth]{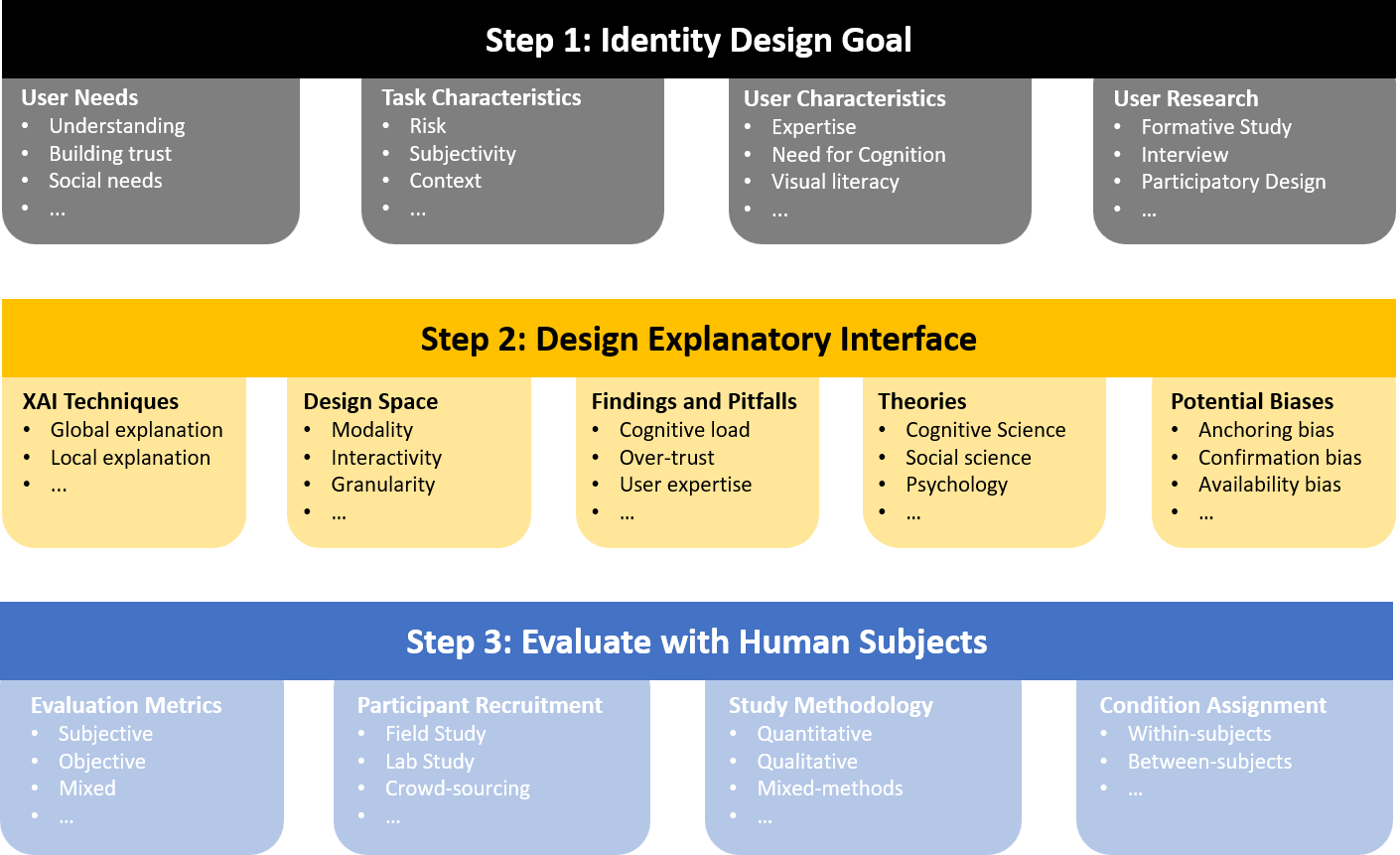}
	\caption{A framework of human-centered XAI design.}
	\label{fig:importance ranking}
\end{figure*}

From the survey papers, we can see a lot of good practices in designing human-centered XAI with empirical studies. Besides their specific design and unique findings, their successful XAI design and user study process are also valuable and insightful for researchers in this area. Thus, based on the lessons learned from the surveyed papers, we propose a framework for the human-centered design of XAI with empirical studies. The framework is illustrated in Figure 1. Overall, there are three stages in the whole procedure, (1) Identify Design Goal, (2) Design Explanatory Interface, (3) Evaluate with Human Subjects.

\subsection{Identify Design Goal}
The first step for human-centered XAI design is to identify the design goal. Generally, the design goal needs to consider user needs and task characteristics. Users' needs for explainability can be quite diverse, such as understanding the AI, building trust, judging the model's fairness, getting informative decision support, etc \cite{mohseni2021multidisciplinary, liao2021human, liao2020questioning}. As described in Section 3, designers can obtain user needs based on user characteristics or user research. There have been some well-established user group categorizations, such as dividing users into AI experts, domain experts, non-experts based on expertise \cite{mohseni2021multidisciplinary}. And designers can adopt a user group-driven explainability needs finding method, to determine the possible user needs based on which category the target users belong to. For example, if the target users are lay people who use the AI application in their daily lives, their needs might include building trust, understanding the AI, etc. Designers can also adopt a user research-driven explainability needs finding method. Also, they can conduct a formative study with the target populations, interview the actual users, invite stakeholders to participate in the prototype design process, and more. From the user research, designers can get more detailed user needs and nuanced implications for the next design step.

Another factor to be considered for identifying the design goal is the task characteristics. In a recent paper, Lai et al. \cite{lai2021towards} categorize the AI-assisted tasks based on application domains, including Law\&Civic, Medicine\&Healthcare, Finance\&Business, Education, Leisure, Professional, Artificial, Generic, and others. Besides the domain-based categorization, they also highlight four dimensions that are critical to distinguishing AI-assisted tasks: (1) task risk (e.g., high, low, or artificial stakes), (2) required expertise in the task, and (3) decision subjectivity, and (4) AI for emulation vs. discovery. Detailed information can be found in \cite{lai2021towards}. Different tasks will require different design goals. For example, a high-stake task in the medicine domain, such as medical disease diagnosis, may require the XAI design to be responsible within strict regulation. While a low-stake task in leisure, such as music recommendation, may require the XAI design to increase user experience.

We recommend that the designers take both user needs and task characteristics into consideration. They can first leverage the user group-driven needs finding method to obtain some alternative user needs, then based on these alternatives design appropriate user research to find the exact user needs in the specific task.

\subsection{Design Explanatory Interface}
The second step for human-centered XAI design is to design the explanatory interface. To begin with, designers need to choose an appropriate XAI technique (in Section 2) based on the specific AI algorithms used in the task and design goals obtained from the first step. Only determining the XAI technique is not enough as the explanatory interface and interaction design can critically affect users' perceptions and the effects of the explanation. Thus, the next step is to decide the design choice from the wide design space as shown in Section 4. To select a specific type of design, apart from focusing on achieving the design goal, we provide three recommendations.

First, designers should be aware of the common findings and pitfalls of different XAI designs from existing empirical studies. This step is not to say that the existing findings and pitfalls of different designs are necessarily applicable to the current task, but to help designers have a general understanding of the potential impact of different design methods on users, so as to guide them in making more reasonable design. For example, if the designer is to design an explanatory interface for an AI-assisted decision-making task, being able to know that providing explanations from AI might lead to users' over-reliance problems can be a great help for them to make better design choices.

Second, designers can refer to XAI-related theories, such as sociology, cognitive science, psychology, and more. Because XAI itself involves \emph{how users receive explanation information}, \emph{how users process explanation information}, and \emph{how users react to specific explanation}, this series of processes will be affected by people's inherent cognitive processes and mental models. Therefore, knowing the basic theoretical knowledge helps to design interpretable interfaces that can conform to the user's cognitive processes and mental models. In recent years, more and more works have recognized the importance of theory-based XAI design \cite{liao2021human, miller2019explanation, wang2019designing}. For example, Liao et al. \cite{liao2021human} highlight the importance of theoretical analysis of human explanations, cognitive and behavioral processes. Wang et al. \cite{wang2019designing} propose a conceptual framework to help designers to map users' reasoning needs to XAI methods. The framework involves four dimensions that describe how humans reason with explanations, including explanation goals, reasoning process, causal explanation type, and elements in rational choice decisions. Miller et al. \cite{miller2019explanation} summarize four major properties of human explanations from a lens of philosophy, psychology, and cognitive science. They find that explanations are often contrastive, selected, and social, and probabilities or statistical explanation can be ineffective. Designers can draw inspiration from these theories.

Third, designers should keep in mind users' potential cognitive biases, especially when designing explanatory interfaces for decision-making tasks. One direct reason for the cognitive bias is that people often make decisions fast with heuristic, which is best described with the dual-process model \cite{kahneman2011thinking}. Users tend to adopt their System 1 thinking in making a decision, which employs heuristics and shortcuts and could lead to cognitive biases. Thus, designers might need to design the explanatory interface while considering whether it can make users engage analytically with the explanations. For example, Wang et al. \cite{wang2019designing} propose several explanation design methods to mitigate different kinds of cognitive biases, including (1) mitigating representativeness bias by prototyping cases of decision outcomes, (2) mitigating availability bias by showing the prevalence of decision outcomes, (3) mitigating anchoring bias by premortem of decision outcome, (4) mitigating confirmation bias by discouraging backward-driven reasoning, (5) facilitating moderate trust by exposing system state and confidence.

\subsection{Evaluate with Human Subjects}
The third step for human-centered XAI design is to evaluate the design with human subjects. First, designers need to determine the evaluation metrics based on the evaluation goals, e.g., users' mental model, task performance, etc. These metrics can be either subjective or objective or mixed based on the specific evaluation goal and the availability of user data, as shown in Section 5. Apart from the evaluation metrics, designers are required to decide participant recruitment method, determine the study methodology, and design condition assignment.

The participant recruitment method includes lab study, crowd-sourcing, online study, field study, etc. From the surveyed empirical studies, we find that nearly half of them recruit non-expert participants via crowd-sourcing platforms such as Amazon Mechanical Turk. The required expertise of participants can determine the participant recruiting method and number of participants \cite{chromik2020taxonomy}. Recruiting difficulty is likely to increase with the required level of participants' expertise \cite{doshi2017towards}. One can recruit novices in large numbers via crowd-sourcing. In contrast, domain or AI experts are usually harder to recruit. They are often invited to a targeted online study, a lab study, or a field study.

The study methodology usually follows a qualitative, quantitative, or mixed study approach. The choice can be influenced by the evaluation goal and the participant recruitment method. For example, qualitative methods or mixed methods are more common in in-lab studies, while quantitative methods are more often seen in crowd-sourcing studies.

Condition assignment should be decided by target evaluation goals. Between-subjects designs investigate the differences between groups of participants, each usually assigned to one XAI condition. In contrast, within-subject designs study differences within individual participants who are assigned to multiple XAI conditions. Also, note that order effects should be considered in a within-subject design.

These are common factors of a user study in the HCI domain, not limited to the empirical study of XAI, thus in this survey, we do not focus on these aspects. However, they are essential components in the whole framework.

\subsection{Summary}
Designers can use this framework to perform a conceptual analysis of 1) how to identify the design goal based on user needs research, user characteristics, task characteristics, and user research; 2) how to design the explanatory interface by considering XAI techniques, design space, common findings and pitfalls obtained from existing work, theories, and potential biases; 3) how to evaluate with human subjects by selecting evaluation metrics, designing appropriate participant recruitment, determining study approach and condition assignment. Although some aspects of the framework are not covered in this survey, we provide recommended references for interested readers to check. We hope this draft framework can help designers build a comprehensive picture of what the process of a human-centered XAI design might look like.